\documentclass[sigconf]{acmart}

\usepackage[title]{appendix}
\usepackage{amsmath}
\usepackage[inline]{enumitem}
\usepackage[noend]{algorithm2e}
\usepackage{caption}
\usepackage{subcaption}
\usepackage{mathtools}
\usepackage{footmisc}
\usepackage{booktabs}
\usepackage{comment}
\usepackage{balance}

\usepackage{xcolor}
\definecolor{boxgrey}{HTML}{F3F3F3}

\def\threedigits#1{\number#1}

\newcommand{\hlbox}[2]{
  \begin{center}
    \fcolorbox{white}{boxgrey}{
      \parbox{0.9\columnwidth}{\noindent \textbf{#1}. \textit{#2}}
    }
  \end{center}
}
\newlist{enumerateinline}{enumerate*}{1}
\setlist[enumerateinline]{
    label=(\roman*)
    ,productjoin={{, }}
    ,productjoin*={{, and }}
    ,after=\unskip{. }
}
\newlist{enumerateinlineempty}{enumerate*}{1}
\setlist[enumerateinlineempty]{
    label=(\roman*)
    ,productjoin={{; }}
    ,productjoin*={{ and }}
}

\AtBeginDocument{
  \providecommand\BibTeX{{
    \normalfont B\kern-0.5em{\scshape i\kern-0.25em b}\kern-0.8em\TeX}}}

\copyrightyear{2022}
\acmYear{2022}
\setcopyright{acmcopyright}\acmConference[SIGIR '22]{Proceedings of the 45th International ACM SIGIR Conference on Research and Development in Information Retrieval}{July 11--15, 2022}{Madrid, Spain}
\acmBooktitle{Proceedings of the 45th International ACM SIGIR Conference on Research and Development in Information Retrieval (SIGIR '22), July 11--15, 2022, Madrid, Spain}
\acmPrice{15.00}
\acmDOI{10.1145/3477495.3532041}
\acmISBN{978-1-4503-8732-3/22/07}

\acmSubmissionID{174}

\begin{document}
\fancyhead{}

\title[Post Processing Recommender Systems with Knowledge Graphs for Recency, Popularity, and Diversity of Explanations]{Post Processing Recommender Systems with Knowledge Graphs for Recency, Popularity, and Diversity of Explanations}

\author{Giacomo Balloccu}
\orcid{0000-0002-6857-7709}
\affiliation{
  \institution{University of Cagliari}
  \streetaddress{}
  \city{Cagliari}
  \country{Italy}}
\email{giacomo.balloccu@unica.it}

\author{Ludovico Boratto}
\orcid{0000-0002-6053-3015}
\affiliation{
  \institution{University of Cagliari}
  \streetaddress{}
  \city{Cagliari}
  \country{Italy}}
\email{ludovico.boratto@acm.org}

\author{Gianni Fenu}
\orcid{0000-0003-4668-2476}
\affiliation{
  \institution{University of Cagliari}
  \streetaddress{}
  \city{Cagliari}
  \country{Italy}}
\email{fenu@unica.it}

\author{Mirko Marras}
\orcid{0000-0003-1989-6057}
\affiliation{
  \institution{University of Cagliari}
  \streetaddress{}
  \city{Cagliari}
  \country{Italy}}
\email{mirko.marras@acm.org}

\begin{abstract}
Existing explainable recommender systems have mainly modeled relationships between recommended and already experienced products, and shaped explanation types accordingly (e.g., movie ``x'' starred by actress ``y'' recommended to a user because that user watched other movies with ``y'' as an actress). 
However, none of these systems has investigated the extent to which properties of a single explanation (e.g., the recency of interaction with that actress) and of a group of explanations for a recommended list (e.g., the diversity of the explanation types) can influence the perceived explaination quality.
In this paper, we conceptualized three novel properties that model the quality of the explanations (linking interaction recency, shared entity popularity, and explanation type diversity) and proposed re-ranking approaches able to optimize for these properties.
Experiments on two public data sets showed that our approaches can increase explanation quality according to the proposed properties, fairly across demographic groups, while preserving recommendation utility.
The source code and data are available at \url{https://github.com/giacoballoccu/explanation-quality-recsys}. 
\end{abstract}

\ccsdesc[500]{Information systems~Recommender systems}
\ccsdesc[500]{Applied computing~Law, social and behavioral sciences}

\keywords{Recommender Systems, Explainability, Fairness, Knowledge Graphs.}

\maketitle

\section{Introduction}

\vspace{1mm} \noindent{\bf Motivation.} 
Explaining to users {\em why} certain results have been provided to them has become an essential property of modern systems. 
Regulations, such as the European General Data Protection Regulation (GDPR), call for a ``right to explanation'', meaning that, under certain conditions, it is mandatory by law to generate awareness for the users on how a model behaves~\cite{GoodmanF17}. 
Explanations have been also proved to have benefits from a business perspective, by increasing trust in the system, helping the users make a decision faster, and persuading a user to try and buy~\cite{Tintarev2007}.
A notable class of decision-support systems that urges supporting explanations are Recommender Systems (RSs). 
Existing RSs often act as black boxes, not offering the user any justification for the recommendations.
Concerted efforts have been devoted to challenge these black boxes to make recommendation a transparent social process~\cite{XianFMMZ19}.

\vspace{1mm} \noindent{\bf State of the Art.} 
Explainable RSs have been created by augmenting traditional models representing user-product interactions with external knowledge, often modelled as Knowledge Graphs (KGs), about the products and the users~\cite{cao-etal-2018-neural, 10.1145/2926718}. 
To this end, prior work has adopted two classes of approaches, based on regularization terms and paths respectively~\cite{arrieta2019explainable}.  
Regularization-based approaches~\cite{CaoWHHC19, CKE10.1145/2939672.2939673, XianFMMZ19, BordesUGWY13, 10.5555/2886521.2886624} have extended the objective function with a term that serves to implicitly encode high-order relationships between users and products from the KG, during model training. 
Though this family of approaches capitalizes on external knowledge to inform the inner functioning of the model, being a feature relevance explanation, no generation of textual explanations for users is provided. {\color{black}Their feature relevance scores are a source of model interpretability rather than explainability from the user's perspective \cite{electronics8080832}.}

\begin{figure*}
\centering
\includegraphics[width=1.0\textwidth]{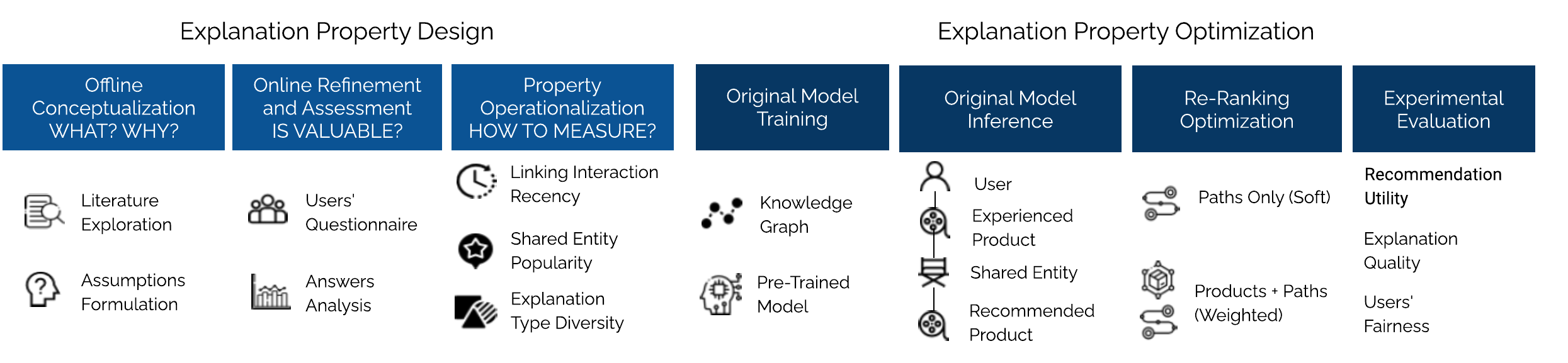}
\vspace{-8mm}
\caption{We adopted a mixed approach combining literature review and user's studies to explore and conceptualize the space of relevant explanation types comprehensively. As a result of this first phase, we identified and operationalized three explanation properties. Recommendations and explainable paths returned by pre-trained models were re-ranked to optimize the explanation properties, and evaluated on recommendation utility, explanation quality, and fairness.} 
\label{fig:path-img}
\end{figure*}

On the other hand, path-based approaches rely on pre-computed paths (tuples) that model high-order relationships between users and products, according to the KG structure~\cite{he2017neural,ripple-net/10.1145/3269206.3271739, Wang_Wang_Xu_He_Cao_Chua_2019, 10.1145/3219819.3219965,10.1109/TKDE.2018.2833443}.
These tuples serve then as an input to the recommendation model during training. 
Compared to regularization-based approaches, this second class includes approaches able to extract one or multiple explainable paths between the recommended item and already experienced products. 
These paths can be in turn translated into textual explanations to be provided to the end users. 
For instance, within the movie domain, a path between a movie already watched by the user ($movie_1$) and a movie recommended to that user ($movie_2$), shaped in the form of $user_1$ \texttt{watched} $movie_1$ \texttt{directed} $director_1$ \texttt{directed} $movie_2$, can be used to provide the textual explanation ``$movie_1$ is recommended to you because you watched another movie directed by $director_1$''.  These paths basically include a past linking interaction ($user_1$ \texttt{watched} $movie_1$), a shared entity ($director_1$), a sharing relationship (\texttt{directed}), and a recommend item ($movie_2$). 

\vspace{1mm} \noindent{\bf Open Issues.} 
Due to the large amount of nodes and arches in the KG, to constrain the path search space, path-based approaches  apply a path selection algorithm~\cite{he2017neural,ripple-net/10.1145/3269206.3271739, Wang_Wang_Xu_He_Cao_Chua_2019, XianFMMZ19} or define a set of meta-path patterns~\cite{10.1145/3219819.3219965,10.1109/TKDE.2018.2833443}. 
{\color{black} However, the path creation and selection is not optimized for the recommendation objective, and the explanation paths are usually pre-computed and then associated to the recommended products. These design choices unfortunately lead to sub-optimal recommendations and explanation paths, with explanations being just post-hoc justifications and not derived from the model inner functioning \cite{4648950}.}  
To overcome this issue, Xian et al.~\cite{XianFMMZ19} proposed to apply reinforcement learning (RL), in the form of policy-guided path reasoning (PGPR), to optimize the recommendation model while searching for paths in the KG. 

Emerging approaches like PGPR rely on a RL agent conditioned on the current user in the KG and trained to navigate to potentially relevant products for that user. 
The agent then samples paths between the user and the recommended products in the KG, based on the probability the agent took that path. 
These paths serve as a basis for the textual explanations that accompany the recommended products. 
However, multiple valid paths between the user and the recommended item exist, and the path leveraged to create the textual explanation is merely selected according to an inner-functioning probability. 
This path selection process does not consider any property connected to how the resulting textual explanation is perceived by the user.  
For instance, a user might prefer explanations linked to more recently experienced products (recency).  
Explainable RSs are hence urging to improve the perceived quality of the explanations from the user's perspective. 

\vspace{1mm} \noindent{\bf Our Approach and Contributions.} 
We believe that the conceptual parts of a path have key characteristics that can significantly influence the perceived explanation quality for the users.
However, there might be a potentially large set of path-related explanation properties to consider. 
Our study therefore explores the space of relevant explanation properties comprehensively through a mixed approach, combining literature review (also including psychological dimensions) and user's studies (investigating which and whether users perceive certain properties as valuable). 
In this way, we identified properties recognized by users as important, such as recency, popularity (extensively connected with novelty and serendipity), and diversity~\cite{DBLP:journals/tiis/KaminskasB17}.     
Considering a single explanation, the \emph{recency} of the past interaction attached to the path  and the \emph{popularity} of the shared entity might influence the perceived quality of the explanations. 
Indeed, more recent linking interactions might lead to a lower memory overload for users (to link back to that past interaction) and explanations better connected to their recent tastes. 
Popular entities might be already known by the user and, possibly, lead to a better understanding of the provided explanation. Conversely, niche entities might help users learn novel links across products in that domain. 
Given that recommendations are often provided as a list, with an explanation for each recommended product, the explanation quality also depends on how explanations are perceived as a whole over the list.  
Their \emph{diversity}, based on the type of relationship in the explanation path (e.g., \texttt{directed} or \texttt{starring}), is a property that can lead to better perceived explanations.

The recognized importance for these properties motivated us to operationalize three novel metrics for recency, popularity, and diversity of explanation. 
We then proposed a suite of re-ranking approaches that optimize the top-$k$ list of recommended products and the accompanying explanations for the proposed metrics.
We assessed the impact of our approaches on recommendation utility and the proposed explanation quality, investigating whether any trade-off aroused. 
Finally, we investigated how these impacts affect different demographic groups protected by law (i.e., gender). 
Figure \ref{fig:path-img} provides a summary of our pipeline. 
Our contribution is threefold:
\begin{enumerate}

\item We define three novel explanation quality metrics to measure (a) the recency of the past linking interaction, (b) the popularity of the shared entity, (c) and the diversity of the explanation types in a recommended list (Section~\ref{sec:exp-properties}).

\item We propose a suite of re-ranking approaches acting on both the original recommended lists and the explainable paths, optimizing for the proposed explanation metrics (Section~\ref{sec:algorithm}). 

\item We evaluate our re-ranking approaches on recommendation utility and explanation metrics, considering both the entire user base and individual demographic groups, on two real-world public data sets, against seven baselines (Section~\ref{sec:experiments}).
\end{enumerate}

\section{Problem Formulation}\label{sec:preliminaries}
We now formally introduce the terminology and notation adopted in our study (Table~\ref{tab:table-of-notation}) and then define the addressed task. 

We denote a \textbf{knowledge graph} as a set of triplets $G = \{(e_h, r, e_t) \\ | \; e_h, e_t \in E, r \in R\}$ where $E$ is the set of entities and $R$ is the set of relationships that might connect two entities. 
Each triplet $(e_h, r, e_t)$ models a relationship $r \in R$ between a head entity $e_h \in E$ and a tail entity $e_t \in E$. 
We assume that at least two classes of entities are present, namely the set $u \in U \subset E$ of users and the set $p \in P \subset E$ of products. 
The relationship $r \in R$ between a user and a product is dependent on the domain (e.g., a user ``watched'' a movie or ``listened to'' a song). 
Example additional entities and relationships might be actors (an actor ``starred'' a movie) or directors (a director ``directed'' a movie) in the movie domain or artists (an artist ``interpreted'' a song) and producers (a producer ``produced'' a song) in the music domain.  
Each relationship $r \in R$ uniquely determines the candidate sets that should be used for the involved head and tail entities (e.g., the actor and movie sets for the relationship ``starred'' or the artist and song sets for the relationship ``interpreted''). 

A \textbf{path} is an alternated sequence of entities and relationships, denoted as $l =\{e_1, r_1, e_2, ..., e_{n-1}, r_n, e_n \}$. 
In the movie domain, an example path between a user and a movie might be $user_1$ \texttt{watched} $movie_1$ \texttt{directed} $director_1$ \texttt{directed} $movie_2$. 
The sequence of relationships included in a path define a path pattern $\pi_l = \{r_1 \circ r_2 \circ ... \circ r_{|\pi_l|} \; | \; r_i \in R, i \leq |\pi_l|\}$.
The pattern of the path in the previous example is \texttt{watched} $\circ$ \texttt{directed} $\circ$ \texttt{directed}. 
We assume that the length of a path $l$ with path pattern $\pi$ is denoted as $|\pi_l|$ (length of 3 for the example path).
The type of a path, denoted as $\omega_l$, is given by the last relationship $r_{|\pi_l|}$ in $\pi_l$ (the type of the example path is ``directed''). 
Being interested in recommending products to users, our focus in this paper is on user-product paths, denoted by $l_{up}$, where $e_1$ is a user and $e_{|\pi_l|}$ is a product. 
We assume that each path $l_{up}$ includes three conceptual parts: a \textit{past interaction} $(e_1 = u \in U, r_1 = r, e_2 = p_1 \in P)$ of a user $u$ with a product $p_1$; 
an \textit{entity chain} $(e_{j-1}, r_{j-1}, e_j)$, with $j=3, \dots, |\pi_l|$, starting from the product $p_1 \in P$ the user interacted with ($e_2 = p_1$) and connecting it with non-product entities $e \notin E_p$; 
a \textit{recommendation} $(e_{|\pi_l|}, r_{|\pi_l|}, e_{|\pi_l| + 1} = p_2 \in P)$ that connects the product $p_2$ to recommend to the rest of the path to user $u$.
Given the guiding example path, ($user_1$ \texttt{watched} $movie_1$) is the past interaction, ($movie_1$ \texttt{directed} $director_1$) is the entity chain, and ($director_1$ \texttt{directed} $movie_2$) is the recommendation. 

\begin{table}[!t]
  \caption{Core notation adopted in our study.}
  \vspace{-4mm}
  \label{tab:table-of-notation}
  \begin{tabular}{ll}
    \toprule
    \textbf{Symbol} & \textbf{Description} \\
    \hline
    \midrule
    $U$ & The set of user entities \\
    $P$ & The set of product entities \\
    $R$ & The user-product relevance matrix\\
    $\Tilde{R}_{up}$ & The predicted user-product relevance score\\
    $\Tilde{P}_u$ & The ordered list of recommended products\\
    \hline 
    $l$ & An alternated path of entities and relationships\\
    $\pi_l$ & The pattern of a path $l$, i.e., list of relationships\\
    $\omega_l$ & The type of a path $l$, i.e., the last relationship in $\pi$\\
    $\hat{l}_{up}$ & The predicted path selected for the explanation\\
    \hline 
    $L_{up}$ & The user-product paths between user and product\\
    $\Tilde{L}_{up}$ & The predicted paths between a user and a product\\
    $\hat{L}_{up}$ & Sorted list of selected paths for a recommended list\\
    $S_l$ & The probability the model took that path \\
    \bottomrule
  \end{tabular}
\end{table}

We assume that users $U$ expressed their \textbf{feedback} for a subset of products in $P$, abstracted to a set of $(u, p)$ pairs implicitly obtained from user activity or $(u, p, v)$ triplets explicitly computed, with $v \in V \subset \mathbb R$. 
Elements in $V$ are either ratings or frequencies (e.g., play counts). 
For the sake of our study, we define the user-product feedback $R \in \mathbb R^{|U| * |P|}$ as a binary matrix, with $R(u, p) = 1$ in case user $u$ interacted with product $p$, $R(u, p) = 0$ otherwise. 
Given this matrix, a recommendation model aims to estimate relevances $\Tilde{R}(u, p) \in [0, 1]$ of unobserved entries in $R$ for a given user and then use them for ranking products. 
This operation can be abstracted as learning a model $\theta : (U, P) \rightarrow \mathbb R$. 
The products are sorted by decreasing relevance for a user, and the top-$k$ products $\Tilde{P}_u$ are recommended.  
For the scope of this paper, we assume that the utility of the recommendations provided by the model for user $u$ is defined as the function $\mathcal C^{RU} : U \xrightarrow[]{} \mathbb{R}$. 

Being focused on improving RS transparency, our goal is not only to recommend a useful set of products, but also to provide relevant explanations as evidence of why a recommendation is provided\footnote{To the best of our knowledge, only path-based explainable RS (e.g., \cite{XianFMMZ19}) are currently able to attach reasoning paths to the recommended products.}.
We assume that $\Tilde{L}_{up} \subset L_{up}$ are the user-products paths predicted for a user $u$ over all paths to a recommended product $p \in \Tilde{P}_u$ in the KG.  
Each path $l \in \Tilde{L}_{up}$ is associated to a score $S_l$ representing the probability that the model used path $l$ to reach $p$ from $u$ over the KG. 
The path $\hat{l}_{up} \in \Tilde{L}_{up}$ with the highest probability $S$ is used to explain the recommendation of $p$ to $u$.

We believe that selecting the explainable path $\hat{l}_{up}$ according to an inner-functioning probability does not necessarily lead to high-quality explanations from the user's perspective.
Therefore, we assume that there exists a set $\mathbb C^{EX}$ of user-related explanation properties denoted with functions in the form $\mathcal C(u) \, : \, (\Tilde{P}_u, \Tilde{L}_{up}) \xrightarrow{} \mathbb N$.
Indeed, an example explanation property to include might be the \emph{recency} of the prior interaction attached to the path of a recommended item. 
Hence, an ideal explainable RS $\theta$ would maximize the expectation on the following objective function:

\vspace{-2mm}
\begin{equation}
\tilde{\theta} = \underset{\theta}{\operatorname{argmax}} \mathop{\mathbb{E}}_{u \; \in \; U} \; C^{RU}(u) + \sum_{\mathcal C \in \; \mathbb C^{EX}} C(u)
\label{eq:problem-definition}
\end{equation}

Considering the possibly heterogeneous nature of the properties in $\mathbb C^{EX}$, we assume that the original recommendation model will be optimized only on recommendation utility ($C^{RU}$), and that the recommended products (and the explainable paths) will be re-ranked so that the explanation properties are maximized, accounting also for recommendation utility. We also assume that the recommendation utility and explanation quality properties are equally weighted for simplicity, leaving user's specific weights as a future work.

\section{Explanation Property Design}\label{sec:exp-properties}
We first present the mixed approach used to explore the space of explanation properties. We combined literature review and user questionnaires to identify, conceptualize, and assess explanation properties users deemed as valuable\footnote{For conciseness, in the paper we only present the main aspects supporting the finally identified three properties from a literature point of view and the extent to which users believe they are important.}. We then turned the identified properties in metrics measurable in a recommended list.  

\subsection{Offline Conceptualization}\label{subsec:conceptualization}
On one hand, we conducted literature reviews to narrow down the range of explanation properties to explore. To the best of our knowledge, no study specifically investigated user-level explanation properties attached to explanation paths returned by an explainable RS. Prior works, such as \cite{FuXGZHGXGSZM20}, explored metrics touching the internal mechanics, e.g., diversity in terms of how many explanation path types are predicted for each recommended item. Further, there might be a potentially large arbitrary set of path-related explanation properties to consider. An approach that includes literature reviews allowed us to have relevant yet grounded examples of explanation properties to start and guide the discussion with users on this topic, being the latter generally expensive in terms of time and effort. 

Our literature analysis covered prior work on the general definition of explainable RSs, e.g., \cite{Tintarev2007}, as well as beyond-accuracy properties investigated in the traditional RS research, e.g., time relevance, diversity, and novelty \cite{DBLP:journals/tiis/KaminskasB17}. We believe that the concepts behind these properties, originally studied with respect to the items being recommended (e.g.,  whether the list of recommendations is diverse and whether it contains time-relevant items) may have a positive impact on the perceived quality of the explanations, if studied with respect to the explanations being provided (e.g., whether the list of explanations accompanying the recommended list is diverse and whether these explanations are linked to items the user recently interacted with). Our analysis led to finally conceptualize three key properties for the produced explanations\footnote{
Explanation quality is a broad topic, so its full assessment would be impossible in the scope of a research study. These properties are by no means exhaustive and other perspectives can be valuable to assess explanation quality.}. We use an example path, namely $u_i$ \texttt{listened} $song_1$ \texttt{featuring} $artist_1$ \texttt{featuring} $song_2$, to showcase each property. 

\vspace{1mm} \noindent{\bf Recency of the Linking Interaction.} 
The first explored property is the recency of the user interaction with the product included in the selected path, i.e., $u_i$ \texttt{listened} $song_1$ in the example path. 
Incorporating the time of interaction into recommendation models is a practice that has shown enhancement in recommendation utility~\cite{LeePP08,DingLO06}, and could influence how users will perceive the final explanation as well.
Indeed, an explanation related to a recent interaction would be intrinsically easier to catch for a user, while older interactions might not be perceived as valuable nor remembered by the users.
As an example, we might consider a highly active user of movie platforms in the past, but was inactive for the last 3 years. 
This user starts to use again the platform, they perform various interactions with new movies, and start receiving new recommendations. 
So, rewarding an explanation based on the freshness of the interaction would be useful. 
Fresher interactions could also be easier to understand yet more timely, compared to those with products associated to very old interactions.

\vspace{1mm} \noindent{\bf Popularity of the Shared Entity.} 
The second explored property is the popularity of the shared entity, i.e., $artist_1$ in the example path. In traditional RS research, popularity is a concept generally connected with novelty (e.g., the less popular the recommended item among other users is, the higher the novelty is) and familiarity (e.g., the more the item is popular among other users, the higher the chance it will be familiar for the user). These two beyond-accuracy properties have been often recognized as important for the recommended items \cite{DBLP:journals/tiis/KaminskasB17}, according to the application scenario. {\color{black} Moreover, the shared entity mentioned in an explanation can act as a source of context, since it can influence the perception of the usefulness of an item \cite{10.1145/2533670.2533675}}. We therefore consider to investigate the extent to which the popularity of the shared entity can influence the perceived quality of the explanation as well. 
For instance, an artist who featured 20 songs might be considered more popular that one who featured 2 songs. 
Indeed, in case a very unpopular recommended product is given, an explanation that contains a popular entity can help the user decide whether that product can be interesting for them. 
In \cite{SwearingenS02}, 70\% of the products that users expressed an interest in buying were familiar products.
These observations are also remarked in~\cite{PuCH12}, which considered the familiarity of the users with the recommended products. 
Conversely, in case the shared entity is too unpopular, the user may not catch the explanation, since they might not know that artist or actor presented in the explanation, or think that is not correlated. 
Providing explanations associated with products that are too popular or redundant could however decrease the filter bubbles in the explanations in the long term~\cite{GedikliJG14}.
Therefore, the popularity of the shared entity might be potentially minimized or maximized according to the overall strategy of the platform, e.g., promoting familiarity or novelty / serendipity. 

\vspace{1mm} \noindent{\bf Diversity of the Explanation Type.} 
The third explanation property is motivated by studies in psychological science, where information diversity is considered a key factor affecting human comprehension and decisions~\cite{AdelmanBQ06}. 
In RS research, diversity is becoming increasingly important, arguing that recommending items by only their predicted relevance increases the risk of producing results that do not satisfy users because the items tend to be too similar to each other~\cite{DBLP:journals/tiis/KaminskasB17}. 
Considering explanations provided in a recommended list as a whole, a possible conceptualization of diversity is that the more explanation path types we present, the better the explanations are perceived. 
For example, in the music domain, we might consider explanation types including \texttt{featured} (as in the example path), \texttt{wrote by}, and \texttt{composed by}, and aim to cover them in the provided explanations in a reasonably balanced way. 
Explanation diversity can help countering the dominance of collaborative-based explanations, in the form ``... because a user who watched your recommended movie has also watched movies you know''. 
This type of explanation might be deemed as too generic - users receiving the recommendation would not know who the other user is, so they cannot trust them. 
Product-based explanations would be more tangible and based on knowledge. 

\subsection{Online Refinement and Assessment}\label{subsec:assessment}
Throughout our process, the above three explanation properties emerged as relevant yet grounded examples to guide the user studies. Such user input is essential to explore the space of explanation types, e.g., refining the conceptualization of the guiding example properties, extending them, or proposing properties not yet covered. Users' feedback also confirms that the envisioned explanation properties are relevant for them, and makes sure that our work is not based on individual examples of explanation properties deemed as important based on arbitrary authors' assumptions. 

\vspace{1mm} \noindent{\bf Design.}  
To ensure the process was scalable, we prepared and sent out a five-minute questionnaire\footnote{A copy of the questions included in the provided questionnaire is available at \url{https://tinyurl.com/exp-quality-survey}.} to a pool of users {\color{black} extracted from the general public through mailing lists and direct messages in social platforms}, covering different demographic groups (e.g., gender, age, country)\footnote{While interviews might lead to more detailed feedback, conducting interviews brings several key challenges, e.g., it is time-consuming to run them at scale, and it is challenging to find interviewers with enough training to conduct the interviews properly across large user bases. We leave this line of research as a future work.}.  
Participation was voluntary, and participants were able to withdraw from the study at any point. 
The responses were collected in an anonymous way, and data from this study was coded confidentially. 
The questionnaire included an introduction describing its purpose and one question for each example property we identified.
We specifically investigated whether users prefer to receive explanations connected to (i) recent/old interactions, (ii) popular/unpopular shared entities, and (iii) a wide/tiny variety of types.
We also included questions to investigate the extent to which, for the users, explanations optimized for these properties promote the transparency, scrutability, effectiveness, persuasiveness, efficiency and satisfaction with the platform, as proposed by \cite{Tintarev2007}. Finally, the users were able to provide, by text, comments on the refinement of the three explanation properties or on ideas about explanation properties not covered in the survey. In total, 104 people (51.4\% female; 48.6\% male) participated in the study, from 29 countries (from Italy 17.14\% and US 12.85\% to Japan 1.85\%, Austria 1.85\%, and Romania 1.85\%), over a wide age range (avg. 27, std. dev. 7, min. 19, max. 48)\footnote{This study was conducted in full compliance to research ethics norms, and specifically the codes and practices established by the ACM Research Ethics Policy.}. 
All participants declared to have a good familiarity with online recommendations.

\vspace{1mm}\noindent\textbf{Results.} 
In the first question, we provided an example timeline of the user's interactions with the platform, together with two possible explanations for an example recommended product (one involving an interaction happened long time ago, and another one involving a more recent interaction). 
From the results, we observed that $64.6\%$ of the participants preferred to see an explanation involving a product closely experienced in time, $6.8\%$ opted for explanations involving older interactions, and the remaining $28.6\%$ of the participants declared that this property would not be relevant for them.

The second question included two explanations for an example product, with an unpopular and a popular shared entity, respectively. 
The results showed that $63.3\%$ of users expressed an interest toward this property, i.e., $40\%$ of the participants preferred a popular shared entity, while $24.3\%$ preferred an unpopular shared entity.
$35.7\%$ of the participants marked this property as not relevant. 

In the third question, we showed two 10-sized recommended lists, with each recommended product accompanied by an explanation.
The first (second) list presented two (four) different types of explanation.  
From the results, we observed that $70\%$ of the participants were in favor of the recommended list accompanied by highly diverse explanation types. 
Surprisingly, $25.7\%$ of the participants expressed their preference towards a low diversity, {\color{black} aligning with prior work that showed how the propensity to diversity depends on the user's personality \cite{10.1145/2468356.2468505}}.
$4.3\%$ of the participants declared that this property would not be relevant.  

In the last question, we asked users to evaluate the extent to which explanations that embed the three proposed properties can improve recommendations from six perspectives \cite{Tintarev2007}. 
For each aspect, the participants could provide a rate between 1 (strongly disagree) and 5 (strongly agree). 
We observed that the majority of the participants agreed on the importance of the properties for better perceiving recommendations in a platform (rate $\ge 4$), i.e., transparency (74.2\%), scrutability (52.8\%), effectiveness (64.2\%), persuasiveness (62.8\%), efficiency (72.8\%), and satisfaction (60\%). 

\subsection{Operationalization}\label{subsec:operationalization}
Based on our conceptualization (offline with literature analysis and online with user studies), we identified the three mentioned explanation properties. They were then operationalized as follows. 

\vspace{1mm}\noindent\textbf{Linking Interaction Recency (LIR).} 
This property serves to quantify the time since the linking interaction in the explanation path occurred. 
Given a user $u \in U$ and the set $P_u$ of products this user interacted with, we denote the list of their interactions, sorted chronologically, by $T_u = [(p^i, t^i)]$, where $p^i \in P_u$ is a product experienced by the user, $t^i \in \mathbb N$ is the timestamp that interaction occurred, and $t^i \leq t^{i+1}$ $\forall i = 1, \dots, |P_u|$. 

We applied an exponentially weighed moving average to the timestamps included in $T_u$, to obtain the LIR of each interaction performed by the user $u$. 
Specifically, given an interaction $(p^i, t^i) \in T_u$, the LIR for that interaction was computed as follows: 

\vspace{-2mm}
\begin{equation} 
LIR(p^i, t^i) = ( 1 - \beta_{LIR} ) \cdot LIR(p^{i-1}, t^{i-1}) + \beta_{LIR} \cdot t^i    
\label{eq:lir}
\end{equation}

where $\beta_{LIR} \in [0, 1]$ is a decay associated to the interaction time, and $LIR(p^1, t^1) = t^1$. 
The $LIR$ values were min-max normalized for each user to lay in the range $[0, 1]$, with values close to $0$ ($1$) meaning that the linking interaction is far away (recent) in time. 
The overall LIR for explanations in a recommended list was obtained by averaging the LIR of the linking interactions for the selected explanation path of each recommended product.

\vspace{1mm}\noindent\textbf{Shared Entity Popularity (SEP).} 
This property serves to quantify the extent to which the shared entity included in an explanation-path is popular. 
We assume that the number of relationships a shared entity is involved in the KG is a proxy of its popularity. 
For instance, the popularity of an actor is computed by counting how many movies that actor starred in.
We denote the list of entities of a given type $\lambda$ in the KG, sorted based on their popularity, by $S_\lambda = [(e^i, v^i)]$, where $e^i \in E_\lambda$ is an entity of type $\lambda$, $v^i \in \mathbb N$ is the number of relationships a shared entity is involved in (in-degree), and $v^i \leq v^{i+1}$ $\forall i = 1, \dots, |E_\lambda|$. 
We applied an exponential decay to the popularity scores in $S_\lambda$, to get the SEP of an entity of type $\lambda$, as: 

\vspace{-2mm}
\begin{equation} 
SEP(e^i, v^i) = ( 1 - \beta_{SEP} ) \cdot SEP(e^{i-1}, v^{i-1}) + \beta_{SEP} \cdot v^i    
\label{eq:sep}
\end{equation}

where $\beta_{SEP}$ is a decay related to the popularity, and $SEP(e^1, v^1) = v^1$. 
The $SEP$ values were min-max normalized for each entity type to lay in the range $[0, 1]$, with values close to $0$ ($1$) when the entity has a low (high) popularity\footnote{The shared entity novelty might be obtained as $SEN(e^i, v^i) = 1-SEP(e^i, v^i)$.}. 
The overall SEP for explanations in a recommended list was obtained by averaging the SEP values of the shared entity in the selected path for each recommended product.

\vspace{1mm}\noindent\textbf{Explanation Type Diversity (ETD)}. 
This property serves to quantify how many different types of explanation are accompanying the recommended products. 
Given a top-$k$ list, recommended to a user $u$, whose products are denoted by $\Tilde{P}_u$ and the corresponding selected explanation path are denoted by $\hat{L}_{u}$, we define as $\omega_{\hat{L}_{u}} = \{\omega_l \; | \; l \in \hat{L}_{u}\}$ the set of path types in explanations for the recommended list of user $u$. 
The ETD of user $u$ is computed as the number of unique types in the selected explanations relative to the minimum between the size of the recommended list $k$ and the total number of possible explanation types. 
Specifically:

\vspace{-2mm}
\begin{equation}
ETD(\Tilde{L}_u) = \frac{|\omega_{\hat{L}_{u}}|}{min(k, |\omega_{L}|)}
\label{eq:etd}
\end{equation}

where $L$ is the list all paths between users and products.
ETD values lay in the range $(0, 1]$, with values close to $0$ ($1$) meaning that the recommended list has a low (high) explanation type diversity. 

\section{Explanation Property Optimization}\label{sec:algorithm}
Given that it is generally hard to embed the proposed properties in the internal model learning process, we propose to re-arrange the recommended lists (and the explanations) returned by a recommendation model, a common practice known as re-ranking. 
{\color{black}This strategy might be limited in its impact, since reordering a small set of recommendations (and explanations) might have a less profound effect than optimizing them during the training process.}
On the other hand, it can be applied to the output of any recommendation model {\color{black} able to produce reasoning paths} and can be easily extended to include any new property. 
We specifically, propose two classes of re-ranking approaches. 
The first class, namely {\em soft}, includes approaches that re-rank the explanation paths for each recommended product according to one or more explanation properties, but not the originally recommended products. 
The second class, namely, {\em weighted}, includes approaches that re-rank both the recommended products and the associated explanation paths.
{\color{black} Soft optimizations are more efficient than weighted optimizations and, therefore, might be the unique solution in case of enormous knowledge graphs. We also conjecture that soft optimizations can lead to relevant gains in explanation quality when the original recommended products are accompanied by a larger yet diverse set of explanations paths. However, considering the experimental setting explained later on in this paper and the preliminary experiments we conducted, soft optimizations are likely to have a minimal impact in case of models purely optimized on recommendation utility. In this paper, we will therefore focus on experimenting with weighted optimization. }

\subsection{Soft Optimization}
With soft optimization approaches, our goal is to adjust the original list of explanation paths for a recommended product, such that the selected path better meets the target explanation property(ies).

Let $u$ be the current user, for every product $p \in \Tilde{P}_u$ in the recommended list, we consider the list $\Tilde{L}_{up}$ of the predicted user-product paths produced by the recommendation model $\theta$.
In case we aim to optimize for a property $C$ (e.g., LIR or SEP) associated to the explanation of a specific recommended item, a soft optimization simply requires to compute the value of the target property for each explanation path in the predicted list and then select the explanation path having the highest value for that property.
Certain properties, such as ETD, are instead defined with respect to the explanations provided until that position of the ranking.
They are therefore related to the list of explanations as a whole, and not to a single explanation.  
In this case, the soft optimization formulation changes in the way the properties are computed, denoting as $C_{\Tilde{P}_u^i}$ the value of the property $C$ when the top-$i$ list, with $i \leq k$, includes items $\Tilde{P}_u^i$. 
We might also be interested in optimizing more than one explanation property together, through a soft optimization, such as $\mathbb C = \{ \text{LIR}, \text{SEP} \}$. 
Finding this path for a recommended product $p \in \Tilde{P}_u$ at position $i$ to user $u$ can be generally summarized as: 

\vspace{-2mm}
\begin{equation}
\hat{l}_{up}^i = \underset{l}{\operatorname{argmax}} \mathop{\mathbb{E}}_{l \in \Tilde{L}_{up}} \; \sum_{C \in \mathbb C} C_{\Tilde{P}_u^i}(u)
\label{eq:soft-multiple}
\end{equation}

{\color {black} We indicated $C_{\Tilde{P}_u^i}$ (instead of $C$ only) to keep the formalization as general as possible. For instance, in case both item- and list-level explanation properties (e.g., $\mathbb C = \{ \text{LIR}, \text{ETD} \}$) are considered, item-level explanation properties (LIR in the example $\mathbb C$) would simply ignore the knowledge about the products $\Tilde{P}_u^i$, since their computation does not depend on the products already recommended. }

\subsection{Weighted Optimization}
Compared to soft optimization, this class of approaches can lead to potentially higher gains in explanation quality, according to the proposed properties. 
However, moving recommended products over the list might lead to a trade-off with recommendation utility. 
For each user $u\in U$, our goal is to determine an optimal set $\Tilde{P}_u$ of $k$ products to be recommended to $u$, so that the target property(ies) pursued by the platform is (are) met while preserving recommendation utility. 
We hence capitalize on a \emph{maximum marginal relevance} approach, with the target property(ies) as the support metric(s). 

For each position $i$ of the ranking, for each recommendable product, we compute a weighted sum between (i) the relevance of that product for the user $u$ and (ii) {\color{black} the extent to which the current recommended list to $u$ would meet the target property(ies), if we include that product in the recommendations, obtained through Eq.~\eqref{eq:soft-multiple}.} 
The weight $\alpha$ assigned to the target property(ies) quantifies how important the target property(ies) is (are) with respect to the relevance of that product for the user. 
Once we compute this weighted score for all products, we find the product that achieves the highest weighted score, and we add that product to the recommendations to $u$ at position $i$. 
The same procedure is repeated until position $k$.
The set $\Tilde{P}_u^\ast$ is obtained by solving the following optimization problem: 

\begin{equation}\label{eq:opt_prob}
    \Tilde{P}_u^i = \mathop{\text{argmax}}_{\Tilde{p} \in P - \Tilde{P}_u^{i-1}}\, (1-\alpha) \; \sum_{p \in \{\Tilde{p}\} \cup \Tilde{P}_u^{i-1}} \Tilde{R}_{up} + \alpha \; \mathop{\text{max}}_{l \in \Tilde{L}_{u\Tilde{p}}} \sum_{C \in \mathbb C} C_{\Tilde{P}_u^{i-1}}(u)
\end{equation}

\noindent where the base case is $\Tilde{P}_u^0 = \emptyset$ and $\alpha \in [0,1]$ is a parameter that expresses the trade-off between relevance and the target property. 
With $\alpha=0$, we yield the output of the original model, not taking the target property into account.
With $\alpha=1$, the output of the recommender is discarded, and we focus only on maximizing the target property. 
This greedy approach yields an ordered list of products, and the list at each step is $(1 - 1/e)$ optimal among the lists of equal size. 
This property fits with the real world, where users may see only the first $k$ recommendations, and the others may become visible after scrolling. 

\vspace{1mm} In terms of computational complexity, once explanation quality is introduced as an objective in the recommendation policy, the LIR (SEP) values for each user (item), for all their interactions, should be computed. Computing all LIR values has $\mathcal{O}(|U| * |T_u|)$ complexity depending on the number of users $|U|$ and the number of interactions per user $|T_u|$, with in general $|T_u| < < |U|$. Computing the LIR value for a new user interaction has $O(1)$ complexity. Conversely, computing all SEP values has a complexity in terms of number of entity types $|\lambda|$ and number of entities per entity type $|S_{\lambda}|$, that is $\mathcal{O}(|\lambda| * |S_{\lambda}|)$, in general $|\lambda| < < |S_{\lambda}|$.  The insertion of a new item, interaction, or entity type would require to (re)compute the SEP values for the entity type that item belongs to. We however assume that SEP values will be updated periodically (and not after every insertion), given that the popularity patterns are not expected to change in the short period. Furthermore, the computation for both LIR and SEP can be easily parallelized. Given the above pre-computed matrices, a soft optimization for a user on LIR, SEP, or ETD has a complexity of $\mathcal{O}(k * n\log{}n)$ being $n = |L_{up}|$ the number of predicted paths between user $u$ and product $p$, and $k$ the size of the recommended list. Conversely, a weighted optimization for a user on LIR, SEP, or ETD has complexity $\mathcal{O}(m\log{}m)$, with $m = |L_{u}|$ being the number of predicted paths for user $u$ by the original model. Optimization on ETD does not require any pre-computation. 

\section{Experimental Evaluation}\label{sec:experiments}
In this section, we aim to evaluate the proposed suite of re-ranking approaches, answering to the following key research questions:

\begin{enumerate}[label={\textbf{RQ\protect\threedigits{\theenumi}}}, leftmargin=*]
    \item Do our approaches lead to a trade-off between recommendation utility and explanation quality?
    \item How does the recommendation utility achieved by our approaches compare to that achieved by other models?
    \item How does the explanation quality achieved by our approaches compare to that achieved by other models?
    \item How do our approaches affect (un)fairness for demographic groups in terms of recommendation utility and explanation quality, compared to the original model?
\end{enumerate}

\subsection{Experimental Setup}

\vspace{1mm}\noindent\textbf{Data Sets.} 
Given that our focus is not only on the impact of our approach on the entire user base but also on the demographic disparities in explanation and recommendation utility, several data sets used in explainable RS could not be used since no user's sensitive attributes are reported.
Hence, our experiments were conducted on MovieLens-1M (ML1M) and LastFM-1B (LASTFM), two public data sets that vary in domain, extensiveness, and sparsity (see Table \ref{tab:data-stats}). 

ML1M is a widely adopted data set for movie recommendation. It reports the users' gender (71.7\% males, 28.3\%  females), age ranges (7 ranges, most represented 25-34 with 34.7\%, least represented $\leq 18$ with 3.67\%), and the timestamp a user-movie interaction occurred. 
We adopted the KG generated in \cite{CaoWHHC19}, discarding the movies (and their corresponding interactions) not present in the KG as entities and the relationship types occurring less than 200 times.   

LASTFM is a music listening data set that reports the users' gender (75.4\% males, 24.6\% females), age (avg. 24.75, std. dev. 8.75, min. 0, max. 112), and the timestamp a user-song interaction occurred. 
We adopted the KG generated in \cite{Wang00LC19}, discarding again the songs (and their corresponding interactions) not present in the KG as entities and the relationship types occurring less than 200 times. 
Users not providing all the three sensitive attributes were discarded.  

\vspace{1mm}\noindent\textbf{Data Preparation.} 
For each data set, we first sorted the interactions of each user chronologically. 
We then performed a training-validation-test split with the 70\% oldest interactions in the training set, the subsequent 10\% in the validation set (adopted for hyper-parameter fine tuning), and the 20\% most recent interactions in the test set. 
We assumed that products already seen by the user were not recommended another time. 
The same pre-processed data sets were used to train, optimize, and test each benchmarked model.

\vspace{1mm}\noindent\textbf{Evaluation Metrics.} 
Our evaluation covered recommendation utility, explanation quality, and fairness, computed on top-10 recommended lists ($k=10$) for the sake of conciseness and clarity. 
We assessed recommendation utility through the Normalized Discounted Cumulative Gain (NDCG) \cite{WangWLHL13}, using binary relevance scores and a base-2 logarithm decay.
We assessed explanation quality from three perspectives, namely linking interaction recency (Eq.~\eqref{eq:lir}), shared entity popularity (Eq.~\eqref{eq:sep}), and explanation type diversity (Eq.~\eqref{eq:etd}). 

For each metric, considering data in the test set, we reported the average value across the entire user base or, when assessing fairness, across users belonging to a given demographic group.
Generally, fairness from the perspective of a given metric was assessed by monitoring the (average, in case of multi-class sensitive attributes) difference of that metric value between demographic groups.
This fairness notion is known in the literature as demographic parity.

\begin{table}[!t]
  \caption{Statistics of pre-processed data sets in this study.}
  \label{tab:data-stats}
  \begin{tabular}{lrrr}
    \newline & \newline & \textbf{ML1M} & \textbf{LASTFM} \\
    \hline
    \toprule
    User-Product & \# Users & 6,040 & 15,773 \\
    Information & \# Products & 3,226 & 47,981 \\
    \newline & \# Interactions & 1,000,209 & 3,955,598 \\
    \hline
    \newline & \# Entities & 16,899 & 114,255 \\
    Knowledge & \# Relations & 1.830.766 & 8,501,868 \\
    Graph & \# Triplets & 156,261 & 464,567 \\
    \newline & \# Relations Types & 10 & 9 \\
    \bottomrule
  \end{tabular}
\end{table}

\begin{figure}[!t]
\centering
\includegraphics[width=0.475\textwidth,height=0.55\textheight]{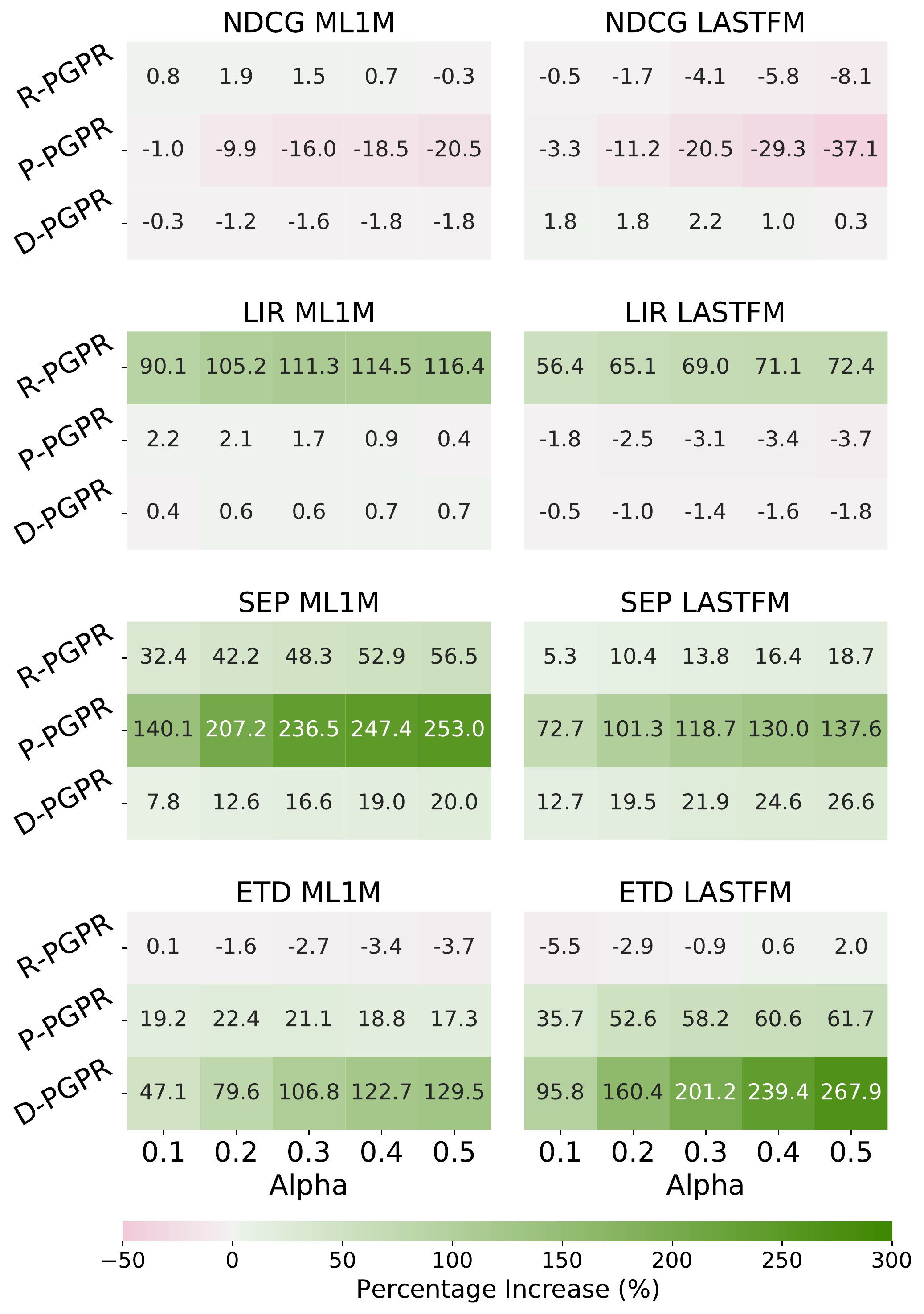}
\caption{{\color{black}Gain / loss of our re-ranking approaches, namely R-PGPR (recency), P-PGPR (popularity), and D-PGPR (diversity), with respect to the original model} in terms of recommendation utility (NDCG) and explanation quality (LIR, SEP, and ETD). We varied $\alpha$ in $[0.1, 0.5]$ and experimented with both ML-1M (left) and LASTFM (right) data sets.}
\label{fig:satt-sopt}
\end{figure}

\vspace{2mm}\noindent\textbf{Benchmarked Models.} 
The considered baselines\footnote{Detailed descriptions are provided in the \href{https://tinyurl.com/tkm87t4s}{repository} and the fine-tuned hyper-parameters are listed in the README of the same \href{https://tinyurl.com/tkm87t4s}{repository}.} included two traditional matrix factorization models (FM \cite{10.1145/2009916.2010002}, NFM \cite{DBLP:conf/uai/RendleFGS09}, BPR \cite{DBLP:conf/uai/RendleFGS09}), three explainable recommendation models based on regularization terms (CFKG \cite{AiACZ18}, CKE \cite{CKE10.1145/2939672.2939673}, KGAT \cite{Wang00LC19}), and one explainable recommendation model based on explanation paths (PGPR \cite{XianFMMZ19}).
Being focused on improving the quality of the textual explanations provided to the users, our suite of re-ranking approaches could be applied only to the output recommendations and explanation paths provided by PGPR, leading to 14 resulting settings.  
We denote those 14 variants as \{R|P|D|RP|RD|PD|RPD\}-PGPR (either soft or weighted), based on what they were optimized for: recency (R : LIR), popularity (P : SEP), diversity (D : ETD) or a combination. 

\hlbox{Remark 1}{{\color{black}To the best of our knowledge, PGPR is the most relevant approach that automatically generates reasoning paths. Defining effective meta-paths requires labor-intensive human domain knowledge and goes beyond the scope of this work.}}

We report results on weighted optimizations and refer the reader to our \href{https://tinyurl.com/tkm87t4s}{repository} for soft optimization results.

\subsection{RQ1: Utility-Explainability Trade-off}\label{sec:tradeoff-rq}
We first investigated whether there exists a trade-off between recommendation utility and the three proposed explanation properties. 
{\color{black} To this end, Figure~\ref{fig:satt-sopt} reports the gain / loss of our proposed re-ranking approaches with respect to the original model, in terms of NDCG, LIR, SEP, and ETD, over $\alpha \in [0.1, 0.5]$. }
Higher values of $\alpha$ led to small gains only ($<5\%$).
We report the results for the weighted approaches optimizing a single property for conciseness.

{\color{black} The heatmaps in the top row of Figure~\ref{fig:satt-sopt} show that optimizing for different metrics influenced recommendation utility differently. In particular, optimizing for LIR (R-PGPR) in ML1M and for ETD (D-PGPR) in LASTFM led to small gains in NDCG (peak at 1.9\% in ML1M with $\alpha = 0.2$ and 2.2\% in LASTFM with $\alpha = 0.3$). In both data sets optimizing for SEP (P-PGPR) resulted in high losses in NDCG with $\alpha > 0.1$.}
{\color{black}The heatmaps in the second row report the results obtained after optimizing for LIR {\color{black} (R-PGPR)}, which resulted in large gains in LIR already at $\alpha = 0.1$ (90.1\% ML1M; 56.4\% LASTFM), until $\alpha = 0.3$ (111.3\% ML1M; 69\% LASTFM). {\color{black} Optimizing for the other explanation properties (P-PGPR, D-PGPR) affected LIR negligibly.}}
{\color {black} The third row of heatmaps highlights that SEP was improved under all the three re-ranking approaches. In particular, directly optimizing SEP (P-PGPR) led to the highest gains (236.5\% ML1M and 118.7\% LASTFM at $\alpha=0.3$). Optimizing for other explanation properties showed a positive indirect gain in SEP, especially while optimizing for LIR with R-PGPR (48.3\% ML1M and 13.8\% LASTFM gain in SEP) and optimizing for ETD with D-PGPR (16.6\% ML1M and 21.9\% LASTFM gain in SEP) at $\alpha=0.3$.}
{\color{black} Finally, the heatmaps in the bottom row reveal that optimizing for ETD (D-PGPR) resulted in large gains for that property already at $\alpha = 0.1$ (47.1\% ML1M; 95.8\% LASTFM). In addition, we observed a positive relation between ETD and SEP. Optimizing for SEP (P-PGPR) led to evident gains in ETD (22.4\% ML1M and 52.6\% LASTFM with $\alpha = 0.2$). Minimal changes emerged in ETD while optimizing for LIR (R-PGPR).} 

\hlbox{Observation 1}{Optimizing for explanation property x not only causes gains in x for the resulting explanations, but positively affects other explanation properties too (e.g., optimizing for SEP leads to gains on both SEP and ETD and viceversa, and SEP benefits from optimizing on LIR), showing a positive interdependence. Even with $\alpha = 0.1$, our re-ranking leads to significant gains  ($\ge 50\%$) on the optimized property, without a significant loss (just $\le 1\%$) in recommendation utility.}

\subsection{RQ2: Impact on Recommendation Utility}\label{sec:performance-comparison-rq}
We then investigated how the recommendation utility achieved by our approaches compares to that achieved by other models. We selected the values of $\alpha$ for our approaches, assuming to work in a context where the platform owners are willing to lose 10\% of NDCG at most to increase as much as possible the quality of the explanations, according to the three proposed properties\footnote{While scientists bring forth the discussion about metrics and design models optimized for them, it is up to stakeholders to select the  most suitable trade-off.}.
Table~\ref{tab:baselines-results} collects the NDCGs obtained under this setting.

Except for CKE, the considered baselines achieved a NDCG ranging in [0.332, 0.340] (ML1M) and [0.134, 0.159] (LASTFM).
Optimizing for LIR resulted in small fluctuations in NDCG (+1.9\% NDCG in ML1M; -1.7\% NDCG in LASTFM), compared to the NDCG achieved by the original PGPR. Similar observations can be made for the optimization of SEP and ETD.
Specifically, we observed a loss of -1\% NDCG for SEP and -1.8\% NDCG for ETD in ML1M; a loss of -3.3\% NDCG for SEP and +0.3\% NDCG for ETD was measured in LASTFM. 
The optimization of two explanation properties jointly showed patterns similar to those observed for a single-property optimization.  
Specifically, combining SEP with ETD impact the most on NDCG (-6.7\% NDCG in ML1M; -7.6\% NDCG in LASTFM). 
The fluctuations in NDCG are, however, highly reduced for the DR-PGPR (-1.1\% NDCG in ML1M; +0.3\% NDCG in LASTFM).
Considering three explanation properties together resulted in a higher drop in NDCG for LASTFM (-8.57\% NDCG) than ML1M (-6.66\% NDCG), with respect to PGPR.

\begin{table}[!t]
\centering
  \caption{Recommendation utility (NDCG).}
  \label{tab:baselines-results}
  \resizebox{0.9\linewidth}{!}{
  \begin{tabular}{rrr|rrr}
    \toprule
    \multicolumn{3}{c}{\textbf{Baseline $\uparrow$}} & \multicolumn{3}{c}{\textbf{Ours $\uparrow$}}\\
    Model & ML1M & LASTFM & Model & ML1M & LASTFM\\
    \midrule
    FM & 0.32 & \bf{0.15} & R-PGPR & \textbf{0.34} & \underline{0.14} \\
    BPR & \underline{0.33} & 0.13 & P-PGPR & \underline{0.33} & 0.13 \\
    NMF & 0.32 & \underline{0.14} & D-PGPR & 0.32 & \underline{0.14} \\
    CKE & \underline{0.33} & \underline{0.14} & DP-PGPR & 0.31 & 0.13\\
    CFKG & 0.27 & 0.08 & PR-PGPR & 0.32 & 0.13 \\
    KGAT & \underline{0.33} & {\bf0.15} & DR-PGPR & 0.32  & 0.13 \\
    PGPR & \underline{0.33} & \underline{0.14} & DPR-PGPR & 0.31 & 0.13 \\
    \bottomrule
     \multicolumn{6}{l}{For each dataset: best result in \textbf{bold}, second-best result \underline{underlined}.}\tabularnewline
  \end{tabular}
  }
\end{table}

\hlbox{Observation 2}{Recommendations obtained through our re-ranking approaches achieved state-of-the-art NDCGs. In both data sets, all re-ranking approaches achieved a NDCG equal or at most 2 points lower than the non-(path-)explainable baselines. This negligible loss is observed in the cases where diversity is included as a property to optimize. Our study interestingly shows that accounting for beyond-accuracy aspects related to user-level explaination quality often does not lead to losses (when observed, they are negligible) in recommendation utility.}

\subsection{RQ3: Impact on Explanation Quality}\label{sec:exp-rq}
Under the same setting introduced in the last experiment, we then investigated how the explanation quality achieved by our approaches compares to that achieved by other models. 
Table~\ref{tab:results-exp-metrics} reports the LIR, SEP, ETD, EQ (the sum of the three explanation metric scores).
Only PGPR is reported as a baseline (see Remark 1). 

In ML1M, the EQ with respect to the specific optimized property is always higher than the EQ of the original PGPR (R-PGPR : +104.6\% LIR, P-PGPR : +142.3\% SEP; D-PGPR : +141.6\% ETD). 
Except for a few cases, each re-ranking approach led to gain on the other two (non-optimized) explanation properties. 
This gain is generally higher while optimizing for LIR (+42.30\% SEP; -1.35\% ETD) than SEP (+2.32\% LIR; +25.00\% ETD) and ETD (+0.22\% LIR; +26.92\% SEP). 
Compared to the single-property optimization settings, performing a joint optimization of ETD and SEP (DP-PGPR) showed an gain in SEP (+11.1\% SEP w.r.t. P-PGPR) and a decrease in ETD (-24.13\% ETD w.r.t. D-PGPR), while LIR remained stable.
The highest gains in EQ were measured when LIR was one of the two optimized properties (DR-PGPR or PR-PGPR).
Specifically, DR-PGPR (PR-PGPR) showed a gain of +16.05\% (+24.81\%) in EQ with respect to the best single-property optimization setting (R-PGPR). 
When we combined the three properties together, the gain in EQ was substantial with respect to the original PGPR (+109.87\%), the best performing single-property optimization (R-PGPR : +69.13\%), and the best performing two-property optimization (PR-PGPR : +111.11\%).
Combining all of them led to close yet still lower scores with respect to the single-property optimizations.

\begin{table}[!t]
\centering
  \caption{Explanation quality (EQ, LIR, EP, ETD).}
    \label{tab:results-exp-metrics}
    \resizebox{1.0\linewidth}{!}{
    \begin{tabular}{r|rrrr|rrrr}
    \toprule
     & \multicolumn{4}{c}{\textbf{ML1M}} & \multicolumn{4}{c}{\textbf{LASTFM}}\\ 
     & EQ $\uparrow$ & LIR $\uparrow$ & SEP $\uparrow$ & ETD $\uparrow$ & EQ $\uparrow$ & LIR $\uparrow$ & SEP $\uparrow$ & ETD $\uparrow$\\
    \midrule
    PGPR & 0.81 & 0.43 & 0.26 & 0.12 & 1.07 & 0.56 & 0.38 & 0.13\\
    \hline
    R-PGPR & 1.37 & {\bf 0.88} & 0.37 & 0.12 & 1.50 & {\bf 0.93} & 0.43 & 0.14\\ 
    P-PGPR & 1.22 & 0.44 & 0.63 & 0.15 & 1.42 & 0.55 & 0.67 & 0.20\\ 
    D-PGPR & 1.05 & 0.43 & 0.33 & {\bf 0.29} & 1.50 & 0.55 & 0.47 & \bf{0.48}\\
    \hline 
    DP-PGPR & 1.37 & 0.44 & \bf{0.70} & 0.22 & 1.56 & 0.55 & \underline{0.68} & 0.32 \\
    PR-PGPR & {\bf 1.71} & \underline{0.86} & \bf{0.70} & 0.14 & \underline{1.80} & \underline{0.86} & 0.50 & 0.43 \\
    DR-PGPR & 1.59 & 0.84 & 0.46 & {\bf 0.29} & {\bf2.01} & \underline{0.86} & \bf{0.69} & \underline{0.46}\\
    \hline
    DPR-PGPR & \underline{1.70} & 0.79 & \underline{0.67} & \underline{0.23} & 1.76 & 0.83 & 0.63 & 0.29\\
    \bottomrule
     \multicolumn{9}{l}{For each metric: best result in \textbf{bold}, second-best result \underline{underlined}.}\tabularnewline
    \end{tabular}
    }
\end{table}

\begin{table}[!t]
\centering
  \caption{{\color{black}The difference between male and female user groups in terms of average recommendation utility (NDCG) and explanation quality (LIR, SEP and ETD).}}
    \label{tab:results-fair-sex}
    \resizebox{1.0\linewidth}{!}{
        \begin{tabular}{r|rrrr|rrrr}
        \toprule
         & \multicolumn{4}{c}{\textbf{ML1M}} & \multicolumn{4}{c}{\textbf{LASTFM}} \\
         & $\Delta$NDCG $\downarrow$ & $\Delta$LIR $\downarrow$ & $\Delta$SEP $\downarrow$ & $\Delta$ETD $\downarrow$ & $\Delta$NDCG $\downarrow$ & $\Delta$LIR $\downarrow$ & $\Delta$SEP $\downarrow$ & $\Delta$ETD $\downarrow$\\
        \midrule
        PGPR & 0.034* & -0.011 & 0.015* & 0.002 & -0.001 & -0.014* & 0.019* & 0.002*\\
        \hline
        R-PGPR & 0.033* & -0.007 & 0.008 & 0.002 & -0.002 & -0.008* & 0.025* & 0.003*\\ 
        P-PGPR & 0.036* & -0.013* & -0.001 & 0.002 & -0.001 & -0.004 & 0.017* & 0.001\\ 
        D-PGPR & 0.034* & -0.014* & 0.006 & -0.011* & 0.009* & -0.010* & 0.013* & -0.015*\\
        \hline 
        DP-PGPR & 0.048* & -0.011* & -0.014* & -0.010* & 0.006* & -0.003 & 0.020* & -0.006*\\ 
        PR-PGPR & 0.034* & -0.006 & -0.009 & 0.002 & -0.003 & -0.008* & 0.020* & -0.004*\\ 
        DR-PGPR & 0.036* & -0.012* & 0.002 & -0.008* & 0.006 & -0.004* & 0.022* & -0.010*\\
        \hline
        DPR-PGPR & 0.041* & -0.006 & -0.009* & -0.005 & 0.005 & -0.006* & 0.024* & -0.004\\
        \bottomrule
        \multicolumn{9}{l}{(*) Statistically significant difference under a non-parametric Kruskal-Wallis test, $p = 0.05$. } \tabularnewline
        \end{tabular}
    }
\end{table}

On LASTFM, we observed patterns similar to ML1M, and therefore we do not discuss them in detail.
The gain in EQ, when we combined the three properties together, was substantial with respect to the original PGPR (+57.60\%), the best performing single-property optimization (D-PGPR : 31.96\%), and the best performing two-property optimization (DR-PGPR : +51.11\%).
Combining all of them led to a SEP higher than that of that setting when SEP is optimized alone; close yet still lower scores were obtained for LIR and ETD with respect to the respective single-property optimization.   

\hlbox{Observation 3}{Gains in explanation quality are large and proportional to the baseline PGPR value, on both data sets.
Higher gains are observed for ETD than other properties. Interestingly, considering all three properties jointly does not lead to the highest overall explanation quality. This highlights possibly diverging optimization patterns across properties that vary according to the domain and characteristics of the data set.}

\subsection{RQ4: Demographic Analysis}\label{sec:fairness-rq}
Our concepts and approaches affect the final recommendation service provided to the users, so investigating their impacts on users is important. We therefore explored demographic differences across benchmarked models in terms of recommendation utility and explanation quality\footnote{Though we would expect to detect unfairness given that no debiasing was carried out, an assessment might uncover key settings where unfairness is more likely to arise and call for a novel class of debiasing that takes into account both user-level explanation and recommendation utility. We left this line of research as a future work.}. 
We focused on demographic groups based on the user's gender (binary according to our data sets) and left experiments on the age attribute on both data sets as a future work. 
{\color{black} Given a certain recommendation property $C$ (utility or explainability), the set of male users $G_1$, and the set of female users $G_2$, we assessed the disparate impact on the demographic groups as follows:}

\begin{equation}
\Delta C (G_1, G_2, C) = \frac{1}{|G_1|} \sum_{u \in G_1} C(u) - \mathop{} \frac{1}{|G_2|} \sum_{u \in G_2} C(u)
\label{eq:gender-diff}
\end{equation}

Table \ref{tab:results-fair-sex} depicts the difference in the average value of either NDCG, LIR, SEP, or ETD, between male and female users, computed as in Eq.~\eqref{eq:gender-diff}. 
In ML1M, PGPR showed statistically significant differences (and therefore unfairness) for both recommendation utility and explanation quality (except for LIR). 
Optimizing for diversity in any setting (single and multiple) led to an increase in unfairness in terms of NDCG and ETD, while the other optimization approaches showed stable or lower unfairness estimates. 
Overall, while still being unfair in NDCG and ETD, our approaches preserve fairness in LIR and (except for R-PGPR) mitigated unfairness in SEP. Similar result patterns were observed in LASTFM.
However, in this case, PGPR showed statistically significant differences (and therefore unfairness) in explanation quality but not in recommendation utility. 
Our approaches again preserved fairness in the latter property (except for DP-PGPR and DR-PGPR), but mitigated unfairness in LIR (except for \{R|PR\}-PGPR). 
Optimizing all the explanation properties jointly, while still being unfair (though less) in SEP and ETD, preserved fairness in NDCG, and mitigated unfairness in LIR. 
{\color{black} We conjecture that the (small) unfairness in utility may be caused by the propagation of the biased patterns in the training set, amplified even more by the embedding creation process and the recommendation pipeline \cite{fairec}.}

\hlbox{Observation 4}{In general, the (un)fairness in recommendation utility measured for the original model is not statistically impacted by our approaches. Similar observations hold for the explanation properties, except for the fact that our approaches mitigate unfairness in ETD in LASTFM. Both the original model and our approaches tend to lead to unfairness in SEP. }

\section{Conclusions and Future Work}
In this paper, we conceptualized, assessed, and operationalized three novel properties to monitor explanation quality at user level in recommendation. We then proposed a suite of re-ranking approaches to optimize for these properties. We evaluated our approaches on two real-world data sets with sensitive attributes showing its effectiveness. Our results showed that not only the proposed approaches improve the overall explanation quality, but also preserve (and sometimes improve) recommendation utility, without exacerbating unfairness across demographic groups, with respect to the original model. {\color{black}From our experiments, it also emerged that there is an interdependence between explanation properties, especially between diversity and popularity.}

Our findings in this study, paired with its limitations, will lead us to extend and operationalize a larger space of user-level explanation properties deemed as important for the next generation of explainable RS. Indeed, the proposed properties are not exhaustive by any means, and further studies will be conducted, through also additional user surveys that explore this domain more comprehensively. For example, a property measuring that explanations are not always based on the same linked interaction can be leveraged to increase diversity. Traditional beyond-accuracy metrics explored in RS research, e.g., serendipity, variety, and novelty, can be further elaborated in the context of explanation paths. Other properties can be also used to control the fair exposure of the entities pertaining to humans (in our case entities like producer, actor, category) in the explanations. Overall, our goal is to improve explanations by  optimizing them for relevant user-level properties. In addition to this, novel approaches that optimize the proposed properties throughout the model learning process will be investigated. Another interesting direction can be to develop explanation subsystems able to turn explanation scores returned by regularized-based explainable RSs into explanation-paths that can be used to provide textual explanations to users. This would also serve to assess the transferability of our re-ranking approaches to more models. On the other hand, assessing generalizability to other domains (e.g., education) will require to extend existing datasets with their KG. Our study also calls for debiasing methods that consider both recommendation utility and explanation quality. In the long term, the impact of the resulting explainable RSs on the platform and its stakeholders will be evaluated via online experiments in the real world.

\begin{acks}
Giacomo Balloccu gratefully acknowledges the Department of Mathematics and Computer Science of the University of Cagliari for the financial support of his PhD scholarship under the ``Analysis and development of methodologies in recommender systems'' project (FONDI RICDIP RECSYS 2019 FENU).
\end{acks}
\balance
\bibliographystyle{ACM-Reference-Format}
\bibliography{sample-sigconf}


\begin{thebibliography}{33}


\ifx \showCODEN    \undefined \def \showCODEN     #1{\unskip}     \fi
\ifx \showDOI      \undefined \def \showDOI       #1{#1}\fi
\ifx \showISBNx    \undefined \def \showISBNx     #1{\unskip}     \fi
\ifx \showISBNxiii \undefined \def \showISBNxiii  #1{\unskip}     \fi
\ifx \showISSN     \undefined \def \showISSN      #1{\unskip}     \fi
\ifx \showLCCN     \undefined \def \showLCCN      #1{\unskip}     \fi
\ifx \shownote     \undefined \def \shownote      #1{#1}          \fi
\ifx \showarticletitle \undefined \def \showarticletitle #1{#1}   \fi
\ifx \showURL      \undefined \def \showURL       {\relax}        \fi
\providecommand\bibfield[2]{#2}
\providecommand\bibinfo[2]{#2}
\providecommand\natexlab[1]{#1}
\providecommand\showeprint[2][]{arXiv:#2}

\bibitem[\protect\citeauthoryear{Adelman, Brown, and Quesada}{Adelman
  et~al\mbox{.}}{2006}]%
        {AdelmanBQ06}
\bibfield{author}{\bibinfo{person}{James~S. Adelman},
  \bibinfo{person}{Gordon~D.A. Brown}, {and} \bibinfo{person}{Jos\'e~F.
  Quesada}.} \bibinfo{year}{2006}\natexlab{}.
\newblock \showarticletitle{Contextual Diversity, Not Word Frequency,
  Determines Word-Naming and Lexical Decision Times}.
\newblock \bibinfo{journal}{\emph{Psychological Science}} \bibinfo{volume}{17},
  \bibinfo{number}{9} (\bibinfo{year}{2006}), \bibinfo{pages}{814--823}.
\newblock
\urldef\tempurl%
\url{https://doi.org/10.1111/j.1467-9280.2006.01787.x}
\showDOI{\tempurl}


\bibitem[\protect\citeauthoryear{Ai, Azizi, Chen, and Zhang}{Ai
  et~al\mbox{.}}{2018}]%
        {AiACZ18}
\bibfield{author}{\bibinfo{person}{Qingyao Ai}, \bibinfo{person}{Vahid Azizi},
  \bibinfo{person}{Xu Chen}, {and} \bibinfo{person}{Yongfeng Zhang}.}
  \bibinfo{year}{2018}\natexlab{}.
\newblock \showarticletitle{Learning Heterogeneous Knowledge Base Embeddings
  for Explainable Recommendation}.
\newblock \bibinfo{journal}{\emph{Algorithms}} \bibinfo{volume}{11},
  \bibinfo{number}{9} (\bibinfo{year}{2018}), \bibinfo{pages}{137}.
\newblock
\urldef\tempurl%
\url{https://doi.org/10.3390/a11090137}
\showDOI{\tempurl}


\bibitem[\protect\citeauthoryear{{Barredo Arrieta}, Díaz-Rodríguez, {Del
  Ser}, Bennetot, Tabik, Barbado, Garcia, Gil-Lopez, Molina, Benjamins,
  Chatila, and Herrera}{{Barredo Arrieta} et~al\mbox{.}}{2020}]%
        {arrieta2019explainable}
\bibfield{author}{\bibinfo{person}{Alejandro {Barredo Arrieta}},
  \bibinfo{person}{Natalia Díaz-Rodríguez}, \bibinfo{person}{Javier {Del
  Ser}}, \bibinfo{person}{Adrien Bennetot}, \bibinfo{person}{Siham Tabik},
  \bibinfo{person}{Alberto Barbado}, \bibinfo{person}{Salvador Garcia},
  \bibinfo{person}{Sergio Gil-Lopez}, \bibinfo{person}{Daniel Molina},
  \bibinfo{person}{Richard Benjamins}, \bibinfo{person}{Raja Chatila}, {and}
  \bibinfo{person}{Francisco Herrera}.} \bibinfo{year}{2020}\natexlab{}.
\newblock \showarticletitle{Explainable Artificial Intelligence (XAI):
  Concepts, taxonomies, opportunities and challenges toward responsible AI}.
\newblock \bibinfo{journal}{\emph{Information Fusion}}  \bibinfo{volume}{58}
  (\bibinfo{year}{2020}), \bibinfo{pages}{82--115}.
\newblock
\showISSN{1566-2535}
\urldef\tempurl%
\url{https://doi.org/10.1016/j.inffus.2019.12.012}
\showDOI{\tempurl}


\bibitem[\protect\citeauthoryear{Bordes, Usunier, Garcia-Dur\'{a}n, Weston, and
  Yakhnenko}{Bordes et~al\mbox{.}}{2013}]%
        {BordesUGWY13}
\bibfield{author}{\bibinfo{person}{Antoine Bordes}, \bibinfo{person}{Nicolas
  Usunier}, \bibinfo{person}{Alberto Garcia-Dur\'{a}n}, \bibinfo{person}{Jason
  Weston}, {and} \bibinfo{person}{Oksana Yakhnenko}.}
  \bibinfo{year}{2013}\natexlab{}.
\newblock \showarticletitle{Translating Embeddings for Modeling
  Multi-Relational Data}. In \bibinfo{booktitle}{\emph{Proceedings of the 26th
  International Conference on Neural Information Processing Systems - Volume
  2}} (Lake Tahoe, Nevada) \emph{(\bibinfo{series}{NIPS'13})}.
  \bibinfo{publisher}{Curran Associates Inc.}, \bibinfo{address}{Red Hook, NY,
  USA}, \bibinfo{pages}{2787–2795}.
\newblock


\bibitem[\protect\citeauthoryear{Cao, Hou, Li, and Liu}{Cao
  et~al\mbox{.}}{2018}]%
        {cao-etal-2018-neural}
\bibfield{author}{\bibinfo{person}{Yixin Cao}, \bibinfo{person}{Lei Hou},
  \bibinfo{person}{Juanzi Li}, {and} \bibinfo{person}{Zhiyuan Liu}.}
  \bibinfo{year}{2018}\natexlab{}.
\newblock \showarticletitle{Neural Collective Entity Linking}. In
  \bibinfo{booktitle}{\emph{Proceedings of the 27th International Conference on
  Computational Linguistics}}. \bibinfo{publisher}{Association for
  Computational Linguistics}, \bibinfo{address}{Santa Fe, New Mexico, USA},
  \bibinfo{pages}{675--686}.
\newblock
\urldef\tempurl%
\url{https://aclanthology.org/C18-1057}
\showURL{%
\tempurl}


\bibitem[\protect\citeauthoryear{Cao, Wang, He, Hu, and Chua}{Cao
  et~al\mbox{.}}{2019}]%
        {CaoWHHC19}
\bibfield{author}{\bibinfo{person}{Yixin Cao}, \bibinfo{person}{Xiang Wang},
  \bibinfo{person}{Xiangnan He}, \bibinfo{person}{Zikun Hu}, {and}
  \bibinfo{person}{Tat{-}Seng Chua}.} \bibinfo{year}{2019}\natexlab{}.
\newblock \showarticletitle{Unifying Knowledge Graph Learning and
  Recommendation: Towards a Better Understanding of User Preferences}. In
  \bibinfo{booktitle}{\emph{The World Wide Web Conference, {WWW} 2019, San
  Francisco, CA, USA, May 13-17, 2019}}. \bibinfo{publisher}{{ACM}},
  \bibinfo{pages}{151--161}.
\newblock
\urldef\tempurl%
\url{https://doi.org/10.1145/3308558.3313705}
\showDOI{\tempurl}


\bibitem[\protect\citeauthoryear{Carvalho, Pereira, and Cardoso}{Carvalho
  et~al\mbox{.}}{2019}]%
        {electronics8080832}
\bibfield{author}{\bibinfo{person}{Diogo~V. Carvalho},
  \bibinfo{person}{Eduardo~M. Pereira}, {and} \bibinfo{person}{Jaime~S.
  Cardoso}.} \bibinfo{year}{2019}\natexlab{}.
\newblock \showarticletitle{Machine Learning Interpretability: A Survey on
  Methods and Metrics}.
\newblock \bibinfo{journal}{\emph{Electronics}} \bibinfo{volume}{8},
  \bibinfo{number}{8} (\bibinfo{year}{2019}).
\newblock
\showISSN{2079-9292}
\urldef\tempurl%
\url{https://doi.org/10.3390/electronics8080832}
\showDOI{\tempurl}


\bibitem[\protect\citeauthoryear{Chen, de~Gemmis, Felfernig, Lops, Ricci, and
  Semeraro}{Chen et~al\mbox{.}}{2013a}]%
        {10.1145/2533670.2533675}
\bibfield{author}{\bibinfo{person}{Li Chen}, \bibinfo{person}{Marco de Gemmis},
  \bibinfo{person}{Alexander Felfernig}, \bibinfo{person}{Pasquale Lops},
  \bibinfo{person}{Francesco Ricci}, {and} \bibinfo{person}{Giovanni
  Semeraro}.} \bibinfo{year}{2013}\natexlab{a}.
\newblock \showarticletitle{Human Decision Making and Recommender Systems}.
\newblock \bibinfo{journal}{\emph{ACM Trans. Interact. Intell. Syst.}}
  \bibinfo{volume}{3}, \bibinfo{number}{3}, Article \bibinfo{articleno}{17}
  (\bibinfo{date}{oct} \bibinfo{year}{2013}), \bibinfo{numpages}{7}~pages.
\newblock
\showISSN{2160-6455}
\urldef\tempurl%
\url{https://doi.org/10.1145/2533670.2533675}
\showDOI{\tempurl}


\bibitem[\protect\citeauthoryear{Chen, Wu, and He}{Chen et~al\mbox{.}}{2013b}]%
        {10.1145/2468356.2468505}
\bibfield{author}{\bibinfo{person}{Li Chen}, \bibinfo{person}{Wen Wu}, {and}
  \bibinfo{person}{Liang He}.} \bibinfo{year}{2013}\natexlab{b}.
\newblock \showarticletitle{How Personality Influences Users' Needs for
  Recommendation Diversity?}. In \bibinfo{booktitle}{\emph{CHI '13 Extended
  Abstracts on Human Factors in Computing Systems}} (Paris, France)
  \emph{(\bibinfo{series}{CHI EA '13})}. \bibinfo{publisher}{Association for
  Computing Machinery}, \bibinfo{address}{New York, NY, USA},
  \bibinfo{pages}{829–834}.
\newblock
\showISBNx{9781450319522}
\urldef\tempurl%
\url{https://doi.org/10.1145/2468356.2468505}
\showDOI{\tempurl}


\bibitem[\protect\citeauthoryear{Ding, Li, and Orlowska}{Ding
  et~al\mbox{.}}{2006}]%
        {DingLO06}
\bibfield{author}{\bibinfo{person}{Yi Ding}, \bibinfo{person}{Xue Li}, {and}
  \bibinfo{person}{Maria~E. Orlowska}.} \bibinfo{year}{2006}\natexlab{}.
\newblock \showarticletitle{Recency-based collaborative filtering}. In
  \bibinfo{booktitle}{\emph{Database Technologies 2006, Proceedings of the 17th
  Australasian Database Conference, {ADC} 2006, Hobart, Tasmania, Australia,
  January 16-19 2006}} \emph{(\bibinfo{series}{{CRPIT}},
  Vol.~\bibinfo{volume}{49})}. \bibinfo{publisher}{Australian Computer
  Society}, \bibinfo{pages}{99--107}.
\newblock
\urldef\tempurl%
\url{https://dl.acm.org/citation.cfm?id=1151747}
\showURL{%
\tempurl}


\bibitem[\protect\citeauthoryear{Edizel, Bonchi, Hajian, Panisson, and
  Tassa}{Edizel et~al\mbox{.}}{2020}]%
        {fairec}
\bibfield{author}{\bibinfo{person}{Bora Edizel}, \bibinfo{person}{Francesco
  Bonchi}, \bibinfo{person}{Sara Hajian}, \bibinfo{person}{Andre Panisson},
  {and} \bibinfo{person}{Tamir Tassa}.} \bibinfo{year}{2020}\natexlab{}.
\newblock \showarticletitle{FaiRecSys: mitigating algorithmic bias in
  recommender systems}.
\newblock \bibinfo{journal}{\emph{International Journal of Data Science and
  Analytics}}  \bibinfo{volume}{9} (\bibinfo{date}{03} \bibinfo{year}{2020}),
  \bibinfo{pages}{197--213}.
\newblock
\urldef\tempurl%
\url{https://doi.org/10.1007/s41060-019-00181-5}
\showDOI{\tempurl}


\bibitem[\protect\citeauthoryear{Fu, Xian, Gao, Zhao, Huang, Ge, Xu, Geng,
  Shah, Zhang, and de~Melo}{Fu et~al\mbox{.}}{2020}]%
        {FuXGZHGXGSZM20}
\bibfield{author}{\bibinfo{person}{Zuohui Fu}, \bibinfo{person}{Yikun Xian},
  \bibinfo{person}{Ruoyuan Gao}, \bibinfo{person}{Jieyu Zhao},
  \bibinfo{person}{Qiaoying Huang}, \bibinfo{person}{Yingqiang Ge},
  \bibinfo{person}{Shuyuan Xu}, \bibinfo{person}{Shijie Geng},
  \bibinfo{person}{Chirag Shah}, \bibinfo{person}{Yongfeng Zhang}, {and}
  \bibinfo{person}{Gerard de Melo}.} \bibinfo{year}{2020}\natexlab{}.
\newblock \showarticletitle{Fairness-Aware Explainable Recommendation over
  Knowledge Graphs}. In \bibinfo{booktitle}{\emph{Proceedings of the 43rd
  International {ACM} {SIGIR} conference on research and development in
  Information Retrieval, {SIGIR} 2020, Virtual Event, China, July 25-30,
  2020}}. \bibinfo{publisher}{{ACM}}, \bibinfo{pages}{69--78}.
\newblock
\urldef\tempurl%
\url{https://doi.org/10.1145/3397271.3401051}
\showDOI{\tempurl}


\bibitem[\protect\citeauthoryear{Gedikli, Jannach, and Ge}{Gedikli
  et~al\mbox{.}}{2014}]%
        {GedikliJG14}
\bibfield{author}{\bibinfo{person}{Fatih Gedikli}, \bibinfo{person}{Dietmar
  Jannach}, {and} \bibinfo{person}{Mouzhi Ge}.}
  \bibinfo{year}{2014}\natexlab{}.
\newblock \showarticletitle{How should {I} explain? {A} comparison of different
  explanation types for recommender systems}.
\newblock \bibinfo{journal}{\emph{Int. J. Hum. Comput. Stud.}}
  \bibinfo{volume}{72}, \bibinfo{number}{4} (\bibinfo{year}{2014}),
  \bibinfo{pages}{367--382}.
\newblock
\urldef\tempurl%
\url{https://doi.org/10.1016/j.ijhcs.2013.12.007}
\showDOI{\tempurl}


\bibitem[\protect\citeauthoryear{Goodman and Flaxman}{Goodman and
  Flaxman}{2017}]%
        {GoodmanF17}
\bibfield{author}{\bibinfo{person}{Bryce Goodman} {and}
  \bibinfo{person}{Seth~R. Flaxman}.} \bibinfo{year}{2017}\natexlab{}.
\newblock \showarticletitle{European Union Regulations on Algorithmic
  Decision-Making and a "Right to Explanation"}.
\newblock \bibinfo{journal}{\emph{{AI} Mag.}} \bibinfo{volume}{38},
  \bibinfo{number}{3} (\bibinfo{year}{2017}), \bibinfo{pages}{50--57}.
\newblock
\urldef\tempurl%
\url{https://doi.org/10.1609/aimag.v38i3.2741}
\showDOI{\tempurl}


\bibitem[\protect\citeauthoryear{He, Liao, Zhang, Nie, Hu, and Chua}{He
  et~al\mbox{.}}{2017}]%
        {he2017neural}
\bibfield{author}{\bibinfo{person}{Xiangnan He}, \bibinfo{person}{Lizi Liao},
  \bibinfo{person}{Hanwang Zhang}, \bibinfo{person}{Liqiang Nie},
  \bibinfo{person}{Xia Hu}, {and} \bibinfo{person}{Tat-Seng Chua}.}
  \bibinfo{year}{2017}\natexlab{}.
\newblock \showarticletitle{Neural Collaborative Filtering}. In
  \bibinfo{booktitle}{\emph{Proceedings of the 26th International Conference on
  World Wide Web}} (Perth, Australia) \emph{(\bibinfo{series}{WWW '17})}.
  \bibinfo{publisher}{International World Wide Web Conferences Steering
  Committee}, \bibinfo{address}{Republic and Canton of Geneva, CHE},
  \bibinfo{pages}{173–182}.
\newblock
\showISBNx{9781450349130}
\urldef\tempurl%
\url{https://doi.org/10.1145/3038912.3052569}
\showDOI{\tempurl}


\bibitem[\protect\citeauthoryear{Hu, Shi, Zhao, and Yu}{Hu
  et~al\mbox{.}}{2018}]%
        {10.1145/3219819.3219965}
\bibfield{author}{\bibinfo{person}{Binbin Hu}, \bibinfo{person}{Chuan Shi},
  \bibinfo{person}{Wayne~Xin Zhao}, {and} \bibinfo{person}{Philip~S. Yu}.}
  \bibinfo{year}{2018}\natexlab{}.
\newblock \showarticletitle{Leveraging Meta-Path Based Context for Top- N
  Recommendation with A Neural Co-Attention Model}. In
  \bibinfo{booktitle}{\emph{Proceedings of the 24th ACM SIGKDD International
  Conference on Knowledge Discovery \& Data Mining}} (London, United Kingdom)
  \emph{(\bibinfo{series}{KDD '18})}. \bibinfo{publisher}{Association for
  Computing Machinery}, \bibinfo{address}{New York, NY, USA},
  \bibinfo{pages}{1531–1540}.
\newblock
\showISBNx{9781450355520}
\urldef\tempurl%
\url{https://doi.org/10.1145/3219819.3219965}
\showDOI{\tempurl}


\bibitem[\protect\citeauthoryear{Kaminskas and Bridge}{Kaminskas and
  Bridge}{2017}]%
        {DBLP:journals/tiis/KaminskasB17}
\bibfield{author}{\bibinfo{person}{Marius Kaminskas} {and}
  \bibinfo{person}{Derek Bridge}.} \bibinfo{year}{2017}\natexlab{}.
\newblock \showarticletitle{Diversity, Serendipity, Novelty, and Coverage: {A}
  Survey and Empirical Analysis of Beyond-Accuracy Objectives in Recommender
  Systems}.
\newblock \bibinfo{journal}{\emph{{ACM} Trans. Interact. Intell. Syst.}}
  \bibinfo{volume}{7}, \bibinfo{number}{1} (\bibinfo{year}{2017}),
  \bibinfo{pages}{2:1--2:42}.
\newblock
\urldef\tempurl%
\url{https://doi.org/10.1145/2926720}
\showDOI{\tempurl}


\bibitem[\protect\citeauthoryear{Lee, Park, and Park}{Lee
  et~al\mbox{.}}{2008}]%
        {LeePP08}
\bibfield{author}{\bibinfo{person}{Tong{-}Queue Lee}, \bibinfo{person}{Young
  Park}, {and} \bibinfo{person}{Yong{-}Tae Park}.}
  \bibinfo{year}{2008}\natexlab{}.
\newblock \showarticletitle{A time-based approach to effective recommender
  systems using implicit feedback}.
\newblock \bibinfo{journal}{\emph{Expert Syst. Appl.}} \bibinfo{volume}{34},
  \bibinfo{number}{4} (\bibinfo{year}{2008}), \bibinfo{pages}{3055--3062}.
\newblock
\urldef\tempurl%
\url{https://doi.org/10.1016/j.eswa.2007.06.031}
\showDOI{\tempurl}


\bibitem[\protect\citeauthoryear{Lin, Liu, Sun, Liu, and Zhu}{Lin
  et~al\mbox{.}}{2015}]%
        {10.5555/2886521.2886624}
\bibfield{author}{\bibinfo{person}{Yankai Lin}, \bibinfo{person}{Zhiyuan Liu},
  \bibinfo{person}{Maosong Sun}, \bibinfo{person}{Yang Liu}, {and}
  \bibinfo{person}{Xuan Zhu}.} \bibinfo{year}{2015}\natexlab{}.
\newblock \showarticletitle{Learning Entity and Relation Embeddings for
  Knowledge Graph Completion}. In \bibinfo{booktitle}{\emph{Proceedings of the
  Twenty-Ninth AAAI Conference on Artificial Intelligence}} (Austin, Texas)
  \emph{(\bibinfo{series}{AAAI'15})}. \bibinfo{publisher}{AAAI Press},
  \bibinfo{pages}{2181–2187}.
\newblock
\showISBNx{0262511290}


\bibitem[\protect\citeauthoryear{Oramas, Ostuni, Noia, Serra, and
  Sciascio}{Oramas et~al\mbox{.}}{2016}]%
        {10.1145/2926718}
\bibfield{author}{\bibinfo{person}{Sergio Oramas},
  \bibinfo{person}{Vito~Claudio Ostuni}, \bibinfo{person}{Tommaso~Di Noia},
  \bibinfo{person}{Xavier Serra}, {and} \bibinfo{person}{Eugenio~Di Sciascio}.}
  \bibinfo{year}{2016}\natexlab{}.
\newblock \showarticletitle{Sound and Music Recommendation with Knowledge
  Graphs}.
\newblock \bibinfo{journal}{\emph{ACM Trans. Intell. Syst. Technol.}}
  \bibinfo{volume}{8}, \bibinfo{number}{2}, Article \bibinfo{articleno}{21}
  (\bibinfo{date}{Oct.} \bibinfo{year}{2016}), \bibinfo{numpages}{21}~pages.
\newblock
\showISSN{2157-6904}
\urldef\tempurl%
\url{https://doi.org/10.1145/2926718}
\showDOI{\tempurl}


\bibitem[\protect\citeauthoryear{Pu, Chen, and Hu}{Pu et~al\mbox{.}}{2012}]%
        {PuCH12}
\bibfield{author}{\bibinfo{person}{Pearl Pu}, \bibinfo{person}{Li Chen}, {and}
  \bibinfo{person}{Rong Hu}.} \bibinfo{year}{2012}\natexlab{}.
\newblock \showarticletitle{Evaluating recommender systems from the user's
  perspective: survey of the state of the art}.
\newblock \bibinfo{journal}{\emph{User Model. User Adapt. Interact.}}
  \bibinfo{volume}{22}, \bibinfo{number}{4-5} (\bibinfo{year}{2012}),
  \bibinfo{pages}{317--355}.
\newblock
\urldef\tempurl%
\url{https://doi.org/10.1007/s11257-011-9115-7}
\showDOI{\tempurl}


\bibitem[\protect\citeauthoryear{Rendle, Freudenthaler, Gantner, and
  Schmidt{-}Thieme}{Rendle et~al\mbox{.}}{2009}]%
        {DBLP:conf/uai/RendleFGS09}
\bibfield{author}{\bibinfo{person}{Steffen Rendle}, \bibinfo{person}{Christoph
  Freudenthaler}, \bibinfo{person}{Zeno Gantner}, {and} \bibinfo{person}{Lars
  Schmidt{-}Thieme}.} \bibinfo{year}{2009}\natexlab{}.
\newblock \showarticletitle{{BPR:} Bayesian Personalized Ranking from Implicit
  Feedback}. In \bibinfo{booktitle}{\emph{{UAI} 2009, Proceedings of the
  Twenty-Fifth Conference on Uncertainty in Artificial Intelligence, Montreal,
  QC, Canada, June 18-21, 2009}}. \bibinfo{publisher}{{AUAI} Press},
  \bibinfo{pages}{452--461}.
\newblock
\urldef\tempurl%
\url{https://dslpitt.org/uai/displayArticleDetails.jsp?mmnu=1\&smnu=2\&article\_id=1630\&proceeding\_id=25}
\showURL{%
\tempurl}


\bibitem[\protect\citeauthoryear{Rendle, Gantner, Freudenthaler, and
  Schmidt-Thieme}{Rendle et~al\mbox{.}}{2011}]%
        {10.1145/2009916.2010002}
\bibfield{author}{\bibinfo{person}{Steffen Rendle}, \bibinfo{person}{Zeno
  Gantner}, \bibinfo{person}{Christoph Freudenthaler}, {and}
  \bibinfo{person}{Lars Schmidt-Thieme}.} \bibinfo{year}{2011}\natexlab{}.
\newblock \showarticletitle{Fast Context-Aware Recommendations with
  Factorization Machines}. In \bibinfo{booktitle}{\emph{Proceedings of the 34th
  International ACM SIGIR Conference on Research and Development in Information
  Retrieval}} (Beijing, China) \emph{(\bibinfo{series}{SIGIR '11})}.
  \bibinfo{publisher}{Association for Computing Machinery},
  \bibinfo{address}{New York, NY, USA}, \bibinfo{pages}{635–644}.
\newblock
\showISBNx{9781450307574}
\urldef\tempurl%
\url{https://doi.org/10.1145/2009916.2010002}
\showDOI{\tempurl}


\bibitem[\protect\citeauthoryear{Shi, Hu, Zhao, and Yu}{Shi
  et~al\mbox{.}}{2019}]%
        {10.1109/TKDE.2018.2833443}
\bibfield{author}{\bibinfo{person}{Chuan Shi}, \bibinfo{person}{Binbin Hu},
  \bibinfo{person}{Wayne~Xin Zhao}, {and} \bibinfo{person}{Philip~S. Yu}.}
  \bibinfo{year}{2019}\natexlab{}.
\newblock \showarticletitle{Heterogeneous Information Network Embedding for
  Recommendation}.
\newblock \bibinfo{journal}{\emph{IEEE Trans. on Knowl. and Data Eng.}}
  \bibinfo{volume}{31}, \bibinfo{number}{2} (\bibinfo{date}{Feb.}
  \bibinfo{year}{2019}), \bibinfo{pages}{357–370}.
\newblock
\showISSN{1041-4347}
\urldef\tempurl%
\url{https://doi.org/10.1109/TKDE.2018.2833443}
\showDOI{\tempurl}


\bibitem[\protect\citeauthoryear{Swearingen and Sinha}{Swearingen and
  Sinha}{2002}]%
        {SwearingenS02}
\bibfield{author}{\bibinfo{person}{Kirsten Swearingen} {and}
  \bibinfo{person}{Rashmi Sinha}.} \bibinfo{year}{2002}\natexlab{}.
\newblock \showarticletitle{Interaction design for recommender systems}. In
  \bibinfo{booktitle}{\emph{Designing Interactive Systems}},
  Vol.~\bibinfo{volume}{6}. Citeseer, \bibinfo{pages}{312--334}.
\newblock


\bibitem[\protect\citeauthoryear{Symeonidis, Nanopoulos, and
  Manolopoulos}{Symeonidis et~al\mbox{.}}{2008}]%
        {4648950}
\bibfield{author}{\bibinfo{person}{Panagiotis Symeonidis},
  \bibinfo{person}{Alexandros Nanopoulos}, {and} \bibinfo{person}{Yannis
  Manolopoulos}.} \bibinfo{year}{2008}\natexlab{}.
\newblock \showarticletitle{Providing Justifications in Recommender Systems}.
\newblock \bibinfo{journal}{\emph{IEEE Transactions on Systems, Man, and
  Cybernetics - Part A: Systems and Humans}} \bibinfo{volume}{38},
  \bibinfo{number}{6} (\bibinfo{year}{2008}), \bibinfo{pages}{1262--1272}.
\newblock
\urldef\tempurl%
\url{https://doi.org/10.1109/TSMCA.2008.2003969}
\showDOI{\tempurl}


\bibitem[\protect\citeauthoryear{Tintarev and Masthoff}{Tintarev and
  Masthoff}{2007}]%
        {Tintarev2007}
\bibfield{author}{\bibinfo{person}{Nava Tintarev} {and} \bibinfo{person}{Judith
  Masthoff}.} \bibinfo{year}{2007}\natexlab{}.
\newblock \showarticletitle{A Survey of Explanations in Recommender Systems}.
\newblock \bibinfo{journal}{\emph{IEEE 23rd International Conference on Data
  Engineering Workshop}}, \bibinfo{pages}{801--810}.
\newblock
\showISBNx{978-1-4244-0832-0}
\urldef\tempurl%
\url{https://doi.org/10.1109/ICDEW.2007.4401070}
\showDOI{\tempurl}


\bibitem[\protect\citeauthoryear{Wang, Zhang, Wang, Zhao, Li, Xie, and
  Guo}{Wang et~al\mbox{.}}{2018}]%
        {ripple-net/10.1145/3269206.3271739}
\bibfield{author}{\bibinfo{person}{Hongwei Wang}, \bibinfo{person}{Fuzheng
  Zhang}, \bibinfo{person}{Jialin Wang}, \bibinfo{person}{Miao Zhao},
  \bibinfo{person}{Wenjie Li}, \bibinfo{person}{Xing Xie}, {and}
  \bibinfo{person}{Minyi Guo}.} \bibinfo{year}{2018}\natexlab{}.
\newblock \showarticletitle{RippleNet: Propagating User Preferences on the
  Knowledge Graph for Recommender Systems}. In
  \bibinfo{booktitle}{\emph{Proceedings of the 27th ACM International
  Conference on Information and Knowledge Management}} (Torino, Italy)
  \emph{(\bibinfo{series}{CIKM '18})}. \bibinfo{publisher}{Association for
  Computing Machinery}, \bibinfo{address}{New York, NY, USA},
  \bibinfo{pages}{417–426}.
\newblock
\showISBNx{9781450360142}
\urldef\tempurl%
\url{https://doi.org/10.1145/3269206.3271739}
\showDOI{\tempurl}


\bibitem[\protect\citeauthoryear{Wang, He, Cao, Liu, and Chua}{Wang
  et~al\mbox{.}}{2019a}]%
        {Wang00LC19}
\bibfield{author}{\bibinfo{person}{Xiang Wang}, \bibinfo{person}{Xiangnan He},
  \bibinfo{person}{Yixin Cao}, \bibinfo{person}{Meng Liu}, {and}
  \bibinfo{person}{Tat{-}Seng Chua}.} \bibinfo{year}{2019}\natexlab{a}.
\newblock \showarticletitle{{KGAT:} Knowledge Graph Attention Network for
  Recommendation}. In \bibinfo{booktitle}{\emph{Proceedings of the 25th {ACM}
  {SIGKDD} International Conference on Knowledge Discovery {\&} Data Mining,
  {KDD} 2019, Anchorage, AK, USA, August 4-8, 2019}}.
  \bibinfo{publisher}{{ACM}}, \bibinfo{pages}{950--958}.
\newblock
\urldef\tempurl%
\url{https://doi.org/10.1145/3292500.3330989}
\showDOI{\tempurl}


\bibitem[\protect\citeauthoryear{Wang, Wang, Xu, He, Cao, and Chua}{Wang
  et~al\mbox{.}}{2019b}]%
        {Wang_Wang_Xu_He_Cao_Chua_2019}
\bibfield{author}{\bibinfo{person}{Xiang Wang}, \bibinfo{person}{Dingxian
  Wang}, \bibinfo{person}{Canran Xu}, \bibinfo{person}{Xiangnan He},
  \bibinfo{person}{Yixin Cao}, {and} \bibinfo{person}{Tat-Seng Chua}.}
  \bibinfo{year}{2019}\natexlab{b}.
\newblock \showarticletitle{Explainable Reasoning over Knowledge Graphs for
  Recommendation}.
\newblock \bibinfo{journal}{\emph{Proceedings of the AAAI Conference on
  Artificial Intelligence}} \bibinfo{volume}{33}, \bibinfo{number}{01}
  (\bibinfo{date}{Jul.} \bibinfo{year}{2019}), \bibinfo{pages}{5329--5336}.
\newblock
\urldef\tempurl%
\url{https://doi.org/10.1609/aaai.v33i01.33015329}
\showDOI{\tempurl}


\bibitem[\protect\citeauthoryear{Wang, Wang, Li, He, and Liu}{Wang
  et~al\mbox{.}}{2013}]%
        {WangWLHL13}
\bibfield{author}{\bibinfo{person}{Yining Wang}, \bibinfo{person}{Liwei Wang},
  \bibinfo{person}{Yuanzhi Li}, \bibinfo{person}{Di He}, {and}
  \bibinfo{person}{Tie{-}Yan Liu}.} \bibinfo{year}{2013}\natexlab{}.
\newblock \showarticletitle{A Theoretical Analysis of {NDCG} Type Ranking
  Measures}. In \bibinfo{booktitle}{\emph{{COLT} 2013 - The 26th Annual
  Conference on Learning Theory, June 12-14, 2013, Princeton University, NJ,
  {USA}}} \emph{(\bibinfo{series}{{JMLR} Workshop and Conference Proceedings},
  Vol.~\bibinfo{volume}{30})}. \bibinfo{publisher}{JMLR.org},
  \bibinfo{pages}{25--54}.
\newblock
\urldef\tempurl%
\url{http://proceedings.mlr.press/v30/Wang13.html}
\showURL{%
\tempurl}


\bibitem[\protect\citeauthoryear{Xian, Fu, Muthukrishnan, de~Melo, and
  Zhang}{Xian et~al\mbox{.}}{2019}]%
        {XianFMMZ19}
\bibfield{author}{\bibinfo{person}{Yikun Xian}, \bibinfo{person}{Zuohui Fu},
  \bibinfo{person}{S. Muthukrishnan}, \bibinfo{person}{Gerard de Melo}, {and}
  \bibinfo{person}{Yongfeng Zhang}.} \bibinfo{year}{2019}\natexlab{}.
\newblock \showarticletitle{Reinforcement Knowledge Graph Reasoning for
  Explainable Recommendation}. In \bibinfo{booktitle}{\emph{Proceedings of the
  42nd International {ACM} {SIGIR} Conference on Research and Development in
  Information Retrieval, {SIGIR} 2019, Paris, France, July 21-25, 2019}}.
  \bibinfo{publisher}{{ACM}}, \bibinfo{pages}{285--294}.
\newblock
\urldef\tempurl%
\url{https://doi.org/10.1145/3331184.3331203}
\showDOI{\tempurl}


\bibitem[\protect\citeauthoryear{Zhang, Yuan, Lian, Xie, and Ma}{Zhang
  et~al\mbox{.}}{2016}]%
        {CKE10.1145/2939672.2939673}
\bibfield{author}{\bibinfo{person}{Fuzheng Zhang},
  \bibinfo{person}{Nicholas~Jing Yuan}, \bibinfo{person}{Defu Lian},
  \bibinfo{person}{Xing Xie}, {and} \bibinfo{person}{Wei-Ying Ma}.}
  \bibinfo{year}{2016}\natexlab{}.
\newblock \showarticletitle{Collaborative Knowledge Base Embedding for
  Recommender Systems}. In \bibinfo{booktitle}{\emph{Proceedings of the 22nd
  ACM SIGKDD International Conference on Knowledge Discovery and Data Mining}}
  (San Francisco, California, USA) \emph{(\bibinfo{series}{KDD '16})}.
  \bibinfo{publisher}{Association for Computing Machinery},
  \bibinfo{address}{New York, NY, USA}, \bibinfo{pages}{353–362}.
\newblock
\showISBNx{9781450342322}
\urldef\tempurl%
\url{https://doi.org/10.1145/2939672.2939673}
\showDOI{\tempurl}


\end{thebibliography}

\end{document}